\documentclass[%
 aip,
 amsmath,amssymb,
 reprint,%
]{revtex4-1}

\usepackage{lipsum}
\usepackage{graphicx}
\usepackage{dcolumn}
\usepackage{bm}
\usepackage{tabularx, multirow, booktabs}

\usepackage[utf8]{inputenc}
\usepackage[T1]{fontenc}
\usepackage{mathptmx}
\newcommand\Alfven{Alfv\'en }

\newcommand{\V}[1]{\mathbf{#1}} 
\newcommand{\T}[1]{{\tt #1}} 

\draft 

\begin{document}

\title[Energization above the Nyquist Frequency]{Observing Particle Energization above the Nyquist Frequency: An Application of the Field-Particle Correlation Technique}

\author{Sarah A. Horvath}
\email[E-mail: ]{sarah-horvath@uiowa.edu}

\author{Gregory G. Howes}
\email[E-mail: ]{gregory-howes@uiowa.edu}

\author{Andrew J. McCubbin}
\email[E-mail:]{andrew-mccubbin@uiowa.edu}

\affiliation{Department of Physics and Astronomy, University of Iowa, Iowa City, Iowa 52242, USA}

\date{\today}

\begin{abstract}
The field-particle correlation technique utilizes single-point measurements to uncover signatures of various particle energization mechanisms in turbulent space plasmas. The signature of Landau damping by electrons has been found in both simulations and \emph{in situ} data from Earth’s magnetosheath using this technique, but instrumental limitations of spacecraft sampling rates present a challenge to discovering the full extent of the presence of Landau damping in the solar wind. Theory predicts that field-particle correlations can recover velocity-space energization signatures even from data that is undersampled with respect to the characteristic frequencies at which the wave damping occurs. To test this hypothesis, we perform a high-resolution gyrokinetic simulation of space plasma turbulence, confirm that it contains signatures of electron Landau damping, and then systematically reduce the time resolution of the data to identify the point at which the signatures become impossible to recover. We find results in support of our theoretical prediction and look for a rule of thumb that can be compared with the measurement capabilities of spacecraft missions to inform the process of applying field-particle correlations to low time resolution data.
\end{abstract}

\maketitle 

\section{Introduction }

Identifying the physical mechanisms which are at work heating particles in the solar wind remains an unconquered challenge in the field of heliophysics. Knowledge of these mechanisms would determine how the solar wind attains a higher temperature profile than can be explained by adiabatic expansion alone, \citep{Richardson:2003} and may provide clues to the resolution of the coronal heating problem. One potential source of particle heating is the dissipation of turbulence. 

Turbulence in the weakly collisional solar wind is composed of nonlinearly interacting \Alfven wave packets \citep{Howes:2013a} that cascade to smaller scales primarily in the direction perpendicular to the mean magnetic field. Despite being nearly collisionless, the turbulence is well modeled by a fluid theory cascade --- with a magnetic energy spectrum having a spectral index between -5/3 (consistent with Goldreich and Schridar's theory of critical balance\citep{Goldreich:1995}) and -3/2 (Bodyrev's theory of dynamic alignment\citep{Boldyrev:2006}) --- until the turbulent fluctuations are on the order of the ion Larmor radius. Near this scale there is a break in the magnetic energy spectrum followed by a steeper power law throughout what is often termed the dissipation range.\citep{Kiyani:2015} Kinetic effects become important at these small scales, including wave-particle interactions that may result in the heating or acceleration of particles, and ultimately in the termination of the turbulent cascade. \citep{Howes:2008b}

Landau damping is one type of collisionless wave-particle interaction that likely is involved in this process. The presence of Landau damping can be determined by patterns in the rate of change of the particle phase-space energy density; specifically, damping via the Landau resonance creates a bipolar energization signature structured around the parallel phase velocity of the damped wave, where parallel and perpendicular are defined relative to the mean magnetic field. This type of velocity-space signature is found using the field-particle correlation (FPC) technique,\citep{Klein:2016,Howes:2017,Klein:2017,Howes:2018} which has been used to characterize a variety of plasma particle energization mechanisms through the unique footprints they create in velocity-space, not only in turbulence, but also in collisionless shocks \citep{Juno:2021} and collisionless magnetic reconnection \citep{McCubbin:2022}.

The field-particle correlation technique operates using only single-point measurements of the plasma distribution function and the electromagnetic fields in order to determine changes in the phase-space energy density ($w_s$) of the particles. The rate of change due to the parallel electric field, which is responsible for Landau damping, is
\begin{equation}
    \left(\frac{\partial w_s}{\partial t}\right)_{E_\parallel} = C_{E_\parallel}.
\end{equation}
Since the relevant observable quantities required to  detect Landau damping are the parallel component electric field $E_\parallel$ and the parallel velocity derivative of the distribution function $\partial f_s/\partial v_\parallel$, the correlation takes the following form:\citep{Klein:2017}
\begin{equation}
C_{E_\parallel}( v_\parallel, v_\perp, t; \tau) = C\left( -q_s \frac{v_\parallel^2}{2} \frac{\partial f_s}{\partial v_\parallel}, E_\parallel \right).
\label{eq:cepar}
\end{equation}

When integrated over all velocity, the correlation is exactly equal to the local rate of change of the particle kinetic energy density, or equivalently to the net rate of electromagnetic work done by the parallel electric field on the particles.\citep{Klein:2017} Importantly, this energy exchange between the particles and the fields includes not only secular gains in particle energy but also conservative oscillations that tend to be larger in amplitude and obscure the secular transfer.\citep{Howes:2017} By choosing a sufficiently long period of time over which to average the correlation, termed the \emph{correlation interval} $\tau$, the oscillatory energy transfer will largely cancel out, revealing the smaller amplitude signature of secular energy transfer. Thus, the field-particle correlation technique is a tool for directly measuring lasting changes in particle phase-space energy density at a single location. The velocity-space structure of these changes in phase-space energy density provides insight into the physical mechanism driving the particle energization.

Signatures of Landau damping have been observed using this technique in both simulations\citep{Klein:2016,Klein:2017,Howes:2018,Klein:2020,Horvath:2020}, spacecraft observations,\citep{Chen:2019, Afshari:2021} and laboratory experiments.\citep{Schroeder:2021} However, finite spacecraft sampling rates put a limit on the study of the highest frequencies in the solar wind. This limit is expressed by the Nyquist-Shannon sampling theorem, which was first implied in the literature by Nyquist in the 1920s\citep{Nyquist:1928} and expressly written by Shannon two decades later.\citep{Shannon:1949} It states that a continuous signal sampled at a rate of $n$ times per second can be completely determined only if it contains no frequencies higher than $n/2$ Hz. From this theorem comes the well known Nyquist frequency ($f_{Ny} = f_s/2$), which is the upper bound on the frequencies that may be analyzed without aliasing from discrete data sampled at a rate $f_s$. 

Despite the unprecedented data collection rates of modern spacecraft, the Nyquist frequency remains low enough to pose a barrier to the analysis of certain wave dynamics. For instance, \textit{Parker Solar Probe} collects electron velocity distributions at a maximum sampling rate of about 4.6~samples per second, \citep{Whittlesey:2020} while waves that may interact with those electrons via resonant processes have frequencies of approximately 30~Hz or higher. This discrepancy between the Nyquist and wave frequencies is sizeable; nonetheless, we predict that this obstacle may be overcome through an extended application of the field-particle correlation technique that recovers signatures of particle energization even when due to interactions with high frequency waves that are undersampled in terms of the Nyquist-Shannon sampling theorem. 

\section{ Science Questions }
\label{sec:Qs}
\begin{figure}[th]
\includegraphics[width=0.48\textwidth, height=0.1875\textwidth]{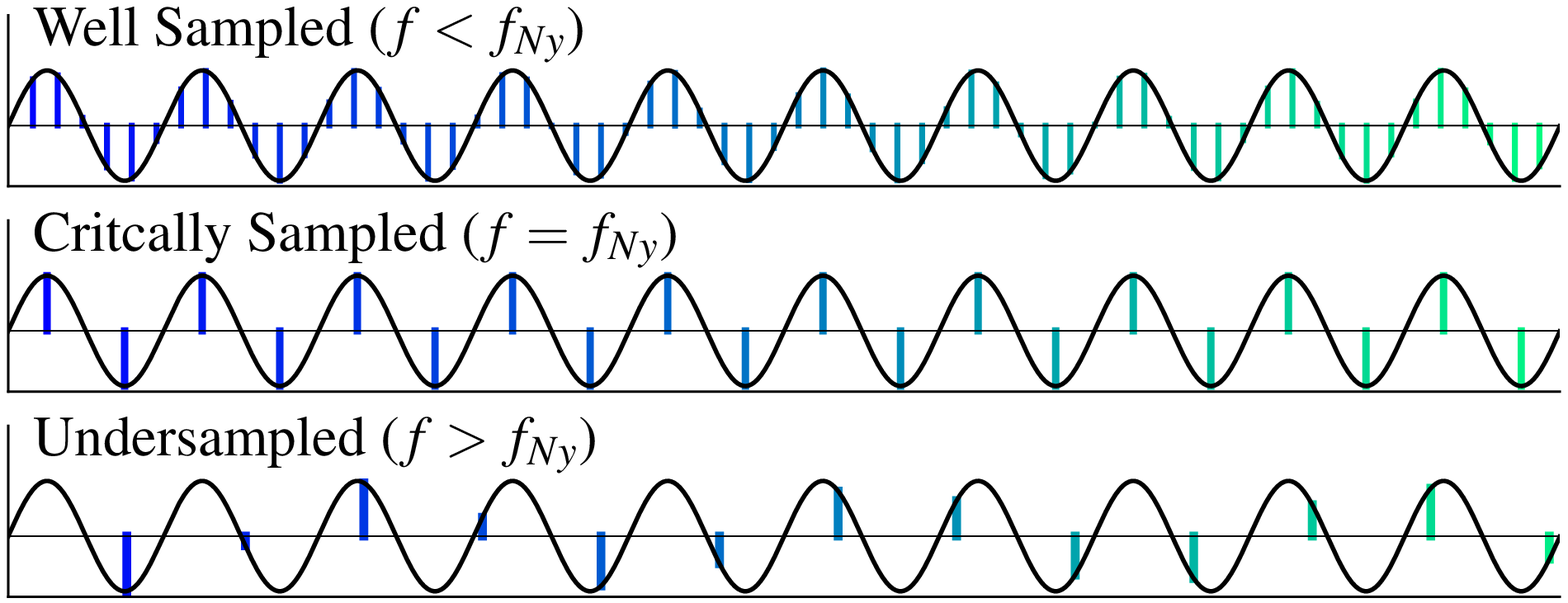}
\includegraphics[width=0.48\textwidth, height=0.141\textwidth]{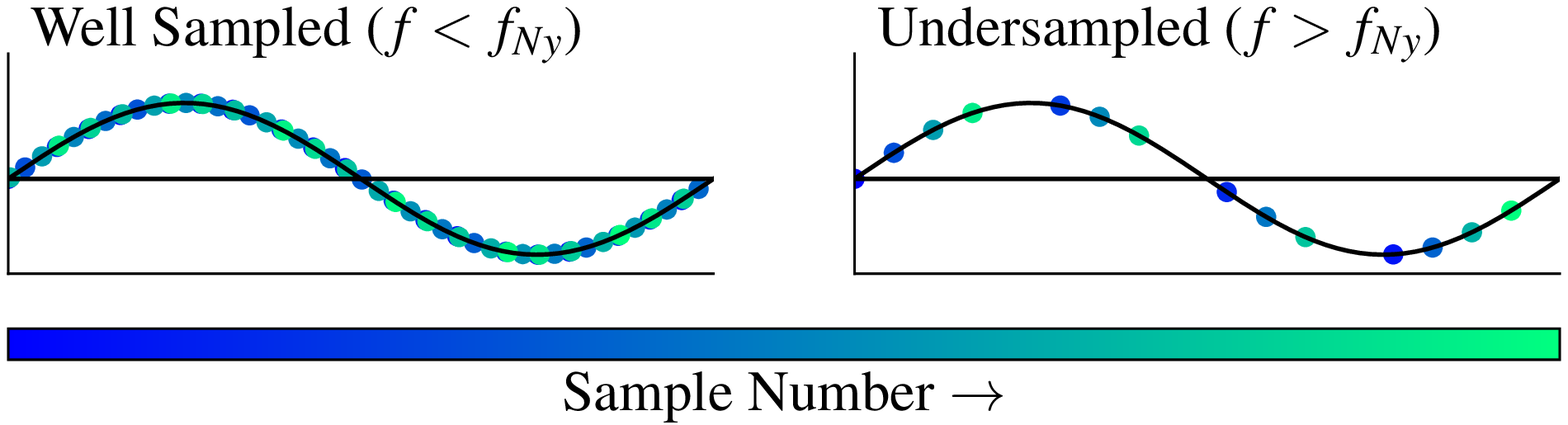}
\caption{\label{fig:f_fNy} \textit{Top} Comparison of a well sampled, critically sampled, and undersampled wave ($f$) as the Nyquist frequency ($f_{Ny} = f_{s}/2$) is reduced. \textit{Bottom} Wave phase at the sampled locations for the well sampled and undersampled cases.} 
\end{figure}

For circumstances when the frequencies involved in resonant wave-particle interactions are higher than the Nyquist frequency of the spacecraft ($f > f_{Ny}$), we hypothesize that velocity-space signatures will still be recoverable via field-particle correlation analysis. Though wave physics cannot be resolved in frequency space above the Nyquist frequency without aliasing, a set of measurements representing all phases of the undersampled frequency can be collected, given enough time. This implies that for a long enough correlation interval, FPC analysis will result in net cancellation of large-amplitude conservative oscillations in energy transfer and reveal signatures of secular energization just as it does for a well sampled frequency. This concept is illustrated in Fig.~\ref{fig:f_fNy}. The same wave (frequency $f$) is sampled at three different rates, resulting in a well sampled ($f < f_{Ny}$), critically sampled ($f = f_{Ny}$), and undersampled ($f > f_{Ny}$) scenario. The bottom panel of Fig.~\ref{fig:f_fNy} shows the wave phase at each sampling time mapped onto a single wave period for the well sampled and undersampled cases. Though the discrete representation of the undersampled wave is incorrect and the true underlying frequency can not be constructed from the samples, a representative set of the wave phases can be collected over a sufficient number of wave periods. 

In this work, we test the prediction that the field-particle correlation technique can be used to recover energization signatures from undersampled plasma data and we look for a rule of thumb to predict the necessary correlation interval given a degree of undersampling. Otherwise stated, our question is: \textit{Can velocity-space signatures of Landau damping be recovered from discretely sampled data when the Nyquist frequency is below the frequency of the damped wave?} An affirmative answer would indicate that the Nyquist frequency is not a barrier to this analysis method for discretely sampled systems, opening the door for the study of very high-frequencey wave-particle energization signatures in data from \emph{Parker Solar Probe} and other missions.

Additionally we ask: \textit{Given an undersampled factor between the wave and Nyquist frequencies ($f/f_{Ny} > 1$), what rule of thumb governs the choice of $\tau$ in the FPC technique in order to reliably recover energization signatures?} Since all information about super-Nyquist frequency waves is unknown in practice -- only their effects on the particles will be returned by this method -- a rule of thumb is sought to guide the choice of correlation interval length for a given spacecraft and plasma environment.

\section{ Methodology }
\subsection{Simulation Setup }
\label{sec:sim}
To test this application of the field-particle correlation technique, we will analyze the energization signatures of electron Landau damping within a high-resolution gyrokinetic simulation. Specifically we use the Astrophysical Gyrokinetics Code (\T{AstroGK}), \citep{Numata:2010} which has successfully modeled heliospheric plasmas in a variety of previous applications.\citep{Howes:2008,Howes:2011,TenBarge:2013b,Howes:2011,Horvath:2020,Schroeder:2021} The simulation is designed with high-resolution in both velocity-space and time to facilitate identification of phase-space energization signatures using field-particle correlations. Then, the dataset is systematically reduced in time resolution and the FPC technique applied again at each step. 

The gyrokinetic simulation depends primarily on three plasma parameters: plasma beta, the ion-to-electron temperature ratio, and the ion-to-electron mass ratio. For this work these are set to: $\beta_i = 0.01$, $T_i/T_e = 1$ and $m_i/m_e = 1836$. The low ratio of thermal to magnetic pressure, realistic mass ratio, and temperature ratio of unity qualitatively match the strongly magnetized environment near the Sun. Note that use of a realistic mass ratio is critical to modeling the finite separation between the ion and electron scale lengths, particularly because the balance of nonlinear energy transfer in the turbulent cascade with the collisionless damping of the turbulent fluctuations depends sensitively on this separation. \citep{Howes:2008,Howes:2011} 

\T{AstroGK} uses the gyro-averaged Maxwell-Boltzmann system of equations to self-consistently evolve a plasma within a 3D Eulerian slab. The simulation domain is elongated in the parallel direction, with dimensions $L_\parallel \times L_\perp^2$ given by $L_\parallel = 2 \pi a_0$ and $L_\perp= \pi \rho_i$. These lengths are normalized by the parameters $a_0$ and $\rho_i$, respectively, where $\rho_i$ is the ion Larmor radius and $a_0$ satisfies the small gyrokinetic expansion parameter: $\rho_i/a_0 \sim \epsilon \ll 1$. \citep{Howes:2006}  The resolution in physical ($x,y,z$) and velocity ($\lambda, \varepsilon$) space is equal to $(n_x, n_y, n_z, n_\lambda, n_\varepsilon, n_s) = (64, 64, 32, 128, 32, 2)$, where $n_s$ is the number of plasma species and the velocity coordinates are given in terms of energy $\varepsilon = v^2$ and pitch angle $\lambda=v_\perp^2/v^2$. Note that though we resolve the fields at this high-resolution in physical space, the velocity distribution function (VDF) is only output for analysis at $24$ individual points dispersed through the box. This emulates \textit{in situ} observations by limiting the analysis to single-point measurements of the VDF as the plasma sweeps past the ``probes.'' 

In our 3D simulation domain of size $L_\parallel \times L_\perp^2$,  a kinetic \Alfven wave with wavevector components $k_{\perp 0}\rho_i  = 2\pi \rho_i /L_\perp = 2$ and $k_{\parallel 0} a_0  = 2\pi a_0/L_\parallel =1 $ has a normalized frequency of $\overline{\omega}_0 \equiv \omega_0/(k_{\parallel} v_A) = 2.044$.  Note that in the MHD limit of $k_\perp \rho_i \ll 1$, the normalized \Alfven wave frequency is simply $\overline{\omega}=1$. We choose to normalize time to the period of this domain-scale kinetic \Alfven wave, $T_0=2\pi /\omega_0$, so that the full duration of the simulation is then $t/T_0 = 68.9$.

Strong turbulence is driven in the simulation via an oscillating Langevin antenna \citep{TenBarge:2014} set to inject four counterpropagating \Alfven waves with $(k_x\rho_i,k_y \rho_i, k_z a_0)=(0,2, \pm 1)$ and $(2,0, \pm 1)$, a driving frequency of $\overline{\omega}_0 = 2.0$, a decorrelation rate of $\gamma_0/\omega_0 = -0.7$, and amplitudes corresponding to critically balanced kinetic \Alfven wave turbulence. \citep{Howes:2008b,Howes:2011b} The nonlinear interaction of the driven waves produces a self-consistent turbulent cascade that is fully resolved in the perpendicular wavenumber range $2 \le k_\perp \rho_i \le 42$.  

As a check that the code ran as expected, we report that the one-dimensional, perpendicular magnetic energy spectrum $E_{B_\perp}(k_\perp) \propto k_\perp^\alpha$ of the simulated plasma is consistent with expectations of the turbulent dissipation range for these plasma conditions, having a spectral index of about $\alpha=-3.2$.  Additionally, we find good energy conservation, with a significant energy gain by the electrons and a negligible gain by the ions. The negligible ion heating is expected for turbulent dissipation at perpendicular scales smaller than the ion Larmor radius, $k_\perp \rho_i > 1$, since the ion contribution to the collisionless damping rate is negligible in this regime (see the lower panel of Fig.~\ref{fig:GKDR}). 

\subsection{ Downsampling the Time Resolution }
\label{sec:time}
{\centering
\begin{table*}[!th]
{\small\centering
\begin{tabular}{ |l|c|c|c|c|c|c|c|c|c|c|c|c| }

\toprule[.1em]
{ $\delta$ } & 0 & 1 & 2 & 3 & 4 & 5 & 6 & 7 & 8 & 9 & 10 & 11 \\
\midrule[.1em]


     { $\Delta t/T_0$ } & 0.017 & 0.034 & 0.067 & 0.134 & 0.268 & 0.536 & 1.070 & 2.137 & 4.287 & 8.599 & 17.099 & 34.594 \\
\midrule[.1em] 

     { $N_{FPC}$ } & 4111 & 2055 & 1027 & 513 & 256 & 128 & 64 & 32 & 16 & 8 & 4 & 2 \\
\midrule[.1em] 

	{  $f_0/f_{Ny}$ }  & 0.034 & 0.067 & 0.134 & 0.268 & 0.534 & 1.06 & 2.11 & 4.14 & 8.04 & 15.0 & 25.7 & 34.6  \\

{ $f_{peak}/f_{Ny}$ }  & 0.092 & 0.184 & 0.368 & 0.737 & 1.48 & 2.96 & 5.91 & 11.8 & 23.6 & 47.3 & 94.6 & 189  \\
\bottomrule[.1em]
\end{tabular}
\caption{Dataset characteristics organized by the downsampling exponent, $\delta$. The average timestep in terms of the domain-scale KAW period is $\Delta t/T_0$; the total number of FPC outputs is $N_{FPC}$; the ratio of the lowest (highest) damped frequency to the Nyquist frequency of the dataset version is $f_0/f_{Ny}$ ($f_{peak}/f_{Ny}$). }\label{tab:res}
}
\end{table*}
}

With this high-time-resolution simulation in hand, next we create downsampled versions of the data. The timestep used in \T{AstroGK} to evolve the gyrokinetic system is too frequent for practical output and much smaller than what is needed for FPC analysis. An FPC diagnostic routine in \T{AstroGK} has been implemented that writes out the information necessary to compute the field-particle correlations---specifically the ion and electron distribution functions and the electromagnetic fields---at a specified set of ``probes'' at fixed spatial positions every $n_{fpc}$ timesteps, where we set $n_{fpc}=2000$ for the simulation analyzed here. In the full dataset, there are $4111$ FPC outputs separated by a sampling interval $\Delta {t}$. The sampling cadence undergoes slight changes during the simulation due to adjustments in the small \T{AstroGK} timestep, which are taken automatically in order to satisfy Courant-Friedrichs-Lewy numerical stability requirements for explicit timestepping of the nonlinear terms in the gyrokinetic equation.\citep{Numata:2010} For the purposes of this study, we choose to define the mean separation between the FPC sampling outputs as the single $\Delta {t}$ for the dataset. Thus the Nyquist frequency is well defined and can be found by the relationship $f_{Ny} = {f}_{s}/2 = 1/(2\Delta {t})$.

The full time series of $4111$ outputs is systematically reduced by factors of two (discarding the final odd data point when necessary) to generate eleven temporally downsampled versions of the simulation data. To produce realistic data, our downsampling process averages together pairs of subsequent outputs in the time series and saves this as a new dataset with corresponding times at the averaged time of each original pair.  Note that the procedure of downsampling in time is applied separately to both the electromagnetic field measurements and the electron velocity distribution at each probe. All twelve versions of the data can be conveniently described by introducing a parameter which we term the \emph{downsampling exponent}, $\delta$, defined such that $2^\delta$ datapoints from the highest resolution dataset have been averaged together to create a given version. For each dataset, the size of the average length between discrete sample times in terms of the driving-scale wave period ($\Delta t /T_0$) is displayed in Table~\ref{tab:res}, along with the total number of FPC outputs contained in that time series, $N_{FPC}$.

\begin{figure}[h!]
\includegraphics[width=0.5\textwidth, height=0.3\textwidth]{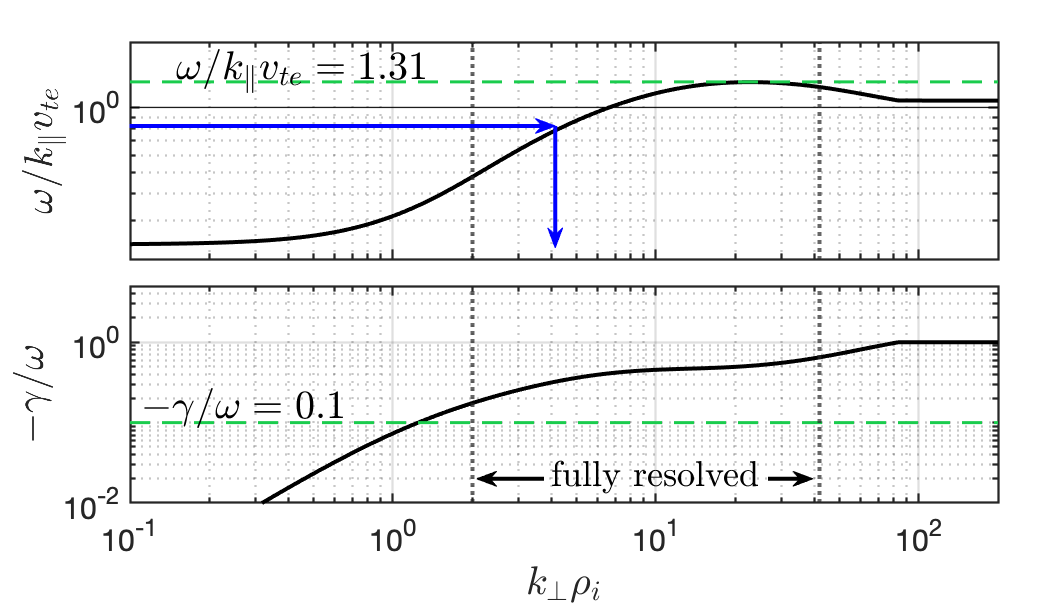}
\caption{\label{fig:GKDR} The linear gyrokinetic dispersion relation for $\beta_i=0.01$, $T_i/T_e = 1$, $m_i/m_e = 1836$, showing plots of both frequency and damping rate. The maximum frequency ($\omega_{max}/k_\parallel v_{te} = 1.31$) and the onset of strong damping ($-\gamma/\omega \ge 0.1$) are marked with horizontal dashed lines. } 
\end{figure} 

The KAWs that compose the turbulent cascade are dispersive for the range of scales and plasma conditions and that we are modeling, as shown by the linear gyrokinetic dispersion relation in Fig.~\ref{fig:GKDR}. This dispersion relation assumes a fully ionized plasma of protons and electrons with Maxwellian equilibrium velocity distributions, plasma parameters $\beta_i=0.01$ and $T_i/T_e = 1$, and a realistic mass ratio $m_i/m_e = 1836$.  Note that the frequency from the linear gyrokinetic dispersion relation, when normalized to $k_\parallel v_A$, is a function of only $\omega/(k_{\parallel} v_A) = \overline{\omega}(k_\perp \rho_i,\beta_i, T_i/T_e)$. \citep{Howes:2006}  To recover the dimensional angular frequency, one may take $\omega= \overline{\omega} k_\parallel v_A$, which demonstrates that the frequency at a fixed value of $k_\perp \rho_i$ is linearly proportional to the parallel wavenumber of the wave.

To explore how a kinetic \Alfven wave interacts collisionlessly with the electron velocity distribution, we instead normalize the parallel phase velocity $\omega/k_\parallel$ to the electron thermal velocity $v_{te} = \sqrt{2T_e/m_e}$, where $T_e$ is expressed in energy units.  In this alternative normalization, we find a minimum frequency of $\omega_0/k_\parallel v_{te} = 0.477$ at the driving scale $k_\perp \rho_i = 2$, which increases monotonically up to a peak of  $\omega_{max}/k_\parallel v_{te} = 1.31$ at a perpendicular wavenumber $k_\perp \rho_i \simeq 22$.

As shown in the lower panel of Fig.~\ref{fig:GKDR}, the entire range of resolved scales $k_\perp \rho_i$ is likely to experience strong collisionless damping by electrons via the Landau resonance, since the normalized damping rate yields values $-\gamma/\omega \ge 0.1$.  Note that for the low $\beta_i=0.01$ plasma conditions, the ion contribution to the damping rate is negligible, so $-\gamma/\omega$ in the range of interest is entirely due to damping by the electrons. The dimensional collisionless damping rate $\gamma = \overline{\gamma} k_\parallel v_A$ is also linearly proportional to the parallel wavenumber, where $\overline{\gamma}$ depends only on the parameters $k_\perp \rho_i$, $\beta_i$, and $T_i/T_e$.  
To determine whether we can recover the velocity-space signature of electron Landau damping above the Nyquist frequency of our sampling, we must estimate the frequency $\omega$ of the wave that is collisionlessly damped. We estimate the frequency of the damped wave by exploiting the fact that the normalized parallel phase velocity in our simulation increases monotonically from $k_\perp \rho_i = 2$ up to the peak at $k_\perp \rho_i \simeq 22$.
When a bipolar velocity-space signature is observed at a given parallel phase velocity $v_\parallel=\omega/k_\parallel$ (determined by the location of the zero crossing in $v_\parallel$), we can use he upper panel of Fig.~\ref{fig:GKDR} to map that normalized parallel phase velocity $\omega/k_\parallel v_{te}$ on the vertical axis to the corresponding 
 value of $k_\perp \rho_i$ on the horizontal axis. \footnote{Above the peak parallel velocity at $k_\perp \rho_i \simeq 22$, the combination of a very large normalized damping rate $-\gamma/\omega >0.6$ and lower amplitudes due to the steep energy spectrum with $\alpha = -3.2$ means that we are unlikely to observe velocity-space signatures of damping at $k_\perp \rho_i > 22$.  Thus, we may connect the parallel phase velocity to the monotonically increasing range of $k_\perp \rho_i$.} 
An example of this determination of $k_\perp \rho_i$ using an observed zero-crossing at a parallel phase velocity $\omega/k_\parallel v_{te}$ is shown by the blue arrows in Fig.~\ref{fig:GKDR}.

The normalized frequency $\overline{\omega}$---or equivalently $\omega/(k_\parallel v_{te}) =\overline{\omega} \beta_i^{-1/2} (T_i/T_e)^{1/2}(m_e/ m_i)^{1/2}$---is a function of only $k_\perp \rho_i$ for fixed  $\beta_i$ and $T_i/T_e$, so the dimensional linear frequency can be calculated from $\overline{\omega}(k_\perp \rho_i)$ using
\begin{equation}
f = \frac{\omega}{2 \pi} = \frac{\overline{\omega}}{2 \pi} k_\parallel v_A
= \frac{\omega}{k_\parallel v_{te}} \beta_i^{1/2} \left(\frac{T_e}{T_i}\right)^{1/2} \left(\frac{m_i}{m_e}\right)^{1/2}\frac{k_\parallel v_A}{2\pi}
\label{eq:freq}
\end{equation}  
Note that the value of $k_\parallel$ for a given damped wave cannot be determined from the single-point information provided by the FPC output. A lower limit of this parallel wavenumber is simply $k_\parallel = k_{\parallel 0}$, corresponding to the assumption of no parallel cascade, so this gives a minimum dimensional frequency for the damped wave.  Scaling theories for the kinetic \Alfven wave cascade at scales $k_\perp \rho_i >1$ predict an expected scaling of the parallel wavenumber as 
$k_\parallel \propto k_\perp^{1/3}$,\citep{TenBarge:2012} although the parallel cascade is expected to weaken or cease once collisionless damping begins to weaken the turbulent cascade at small scales. \citep{Howes:2011b} If this scaling of the parallel cascade indeed were to hold up to the peak of the wave phase velocity at  $k_\perp \rho_i \simeq 22$, this would imply an estimated increased factor of $(22/2)^{1/3} \simeq 2.2$ in the frequency of the damped wave.  In the following analysis, we neglect this potential increase in the frequency by taking $k_\parallel = k_{\parallel 0}$. Thus, we obtain a \emph{lower limit} on the ability of the FPC technique to recover velocity-space signatures of the collsionless damping of waves with frequencies above the Nyquist frequency of the sampling.
 
Therefore, using the parallel phase velocity of the zero-crossings of bipolar velocity-space signatures, we compute the frequency of the damped wave by $f =\overline{\omega} k_{\parallel 0} v_A/2 \pi$.  Dividing this damped wave frequency by the Nyquist frequency of the dataset yields a dimensionless measure of the technique performance for waves above the Nyquist frequency, $f/f_{Ny}$. 
A useful way to characterize the sampling of the kinetic \Alfven waves for a given downsampled dataset is to determine the ratio $f_0/f_{Ny}$ for the minimum wave frequency in the domain corresponding to $\omega/k_\parallel v_{te} = 0.477$ as well as the ratio 
$f_{max}/f_{Ny}$ for the maximum wave frequency corresponding to $\omega/k_\parallel v_{te} = 1.31$.  These characteristic values are listed in Table~\ref{tab:res}, which shows that some waves begin to be undersampled ($f/f_{Ny} > 1$) for downsampling exponent $\delta=4$.

\subsection{ Field-Particle Correlation Analysis }
\label{sec:FPCA}
\begin{figure*}[ht!]
\begin{centering}
\hfill{
\includegraphics[width=0.32\textwidth, height=0.25\textwidth]{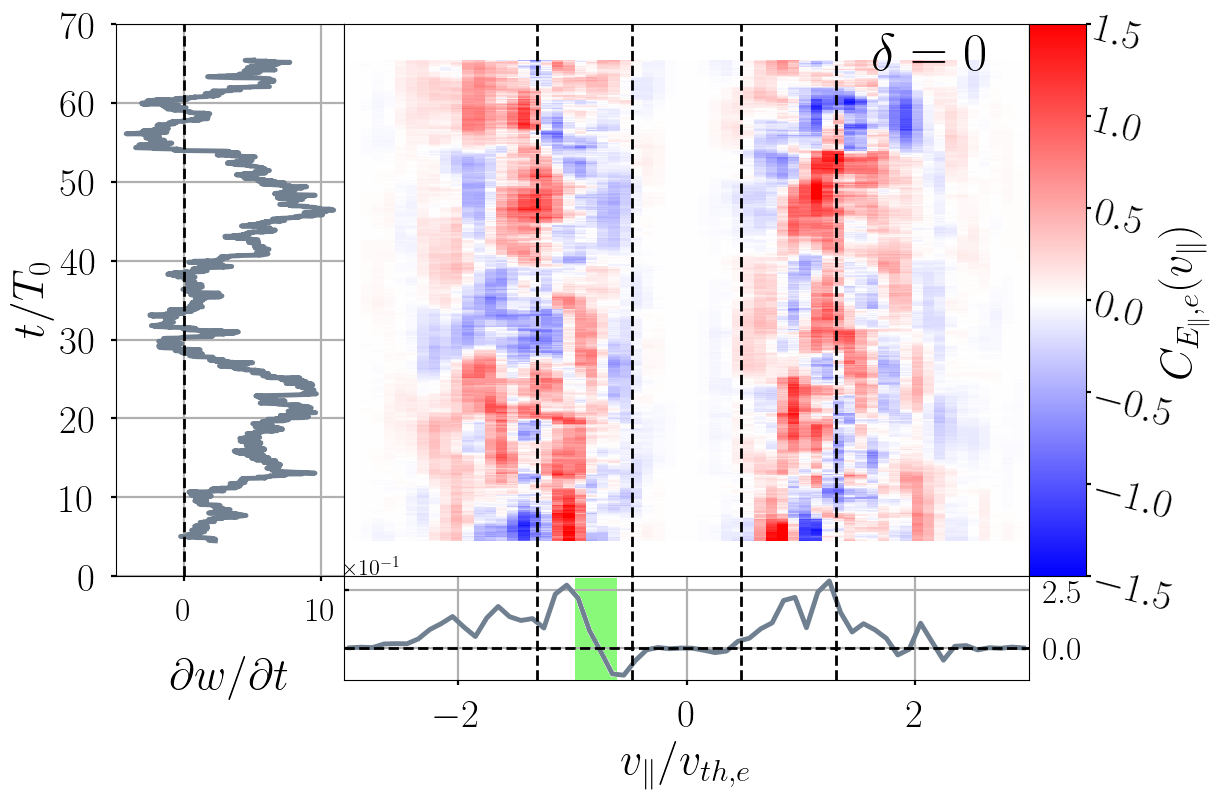}
\includegraphics[width=0.32\textwidth, height=0.25\textwidth]{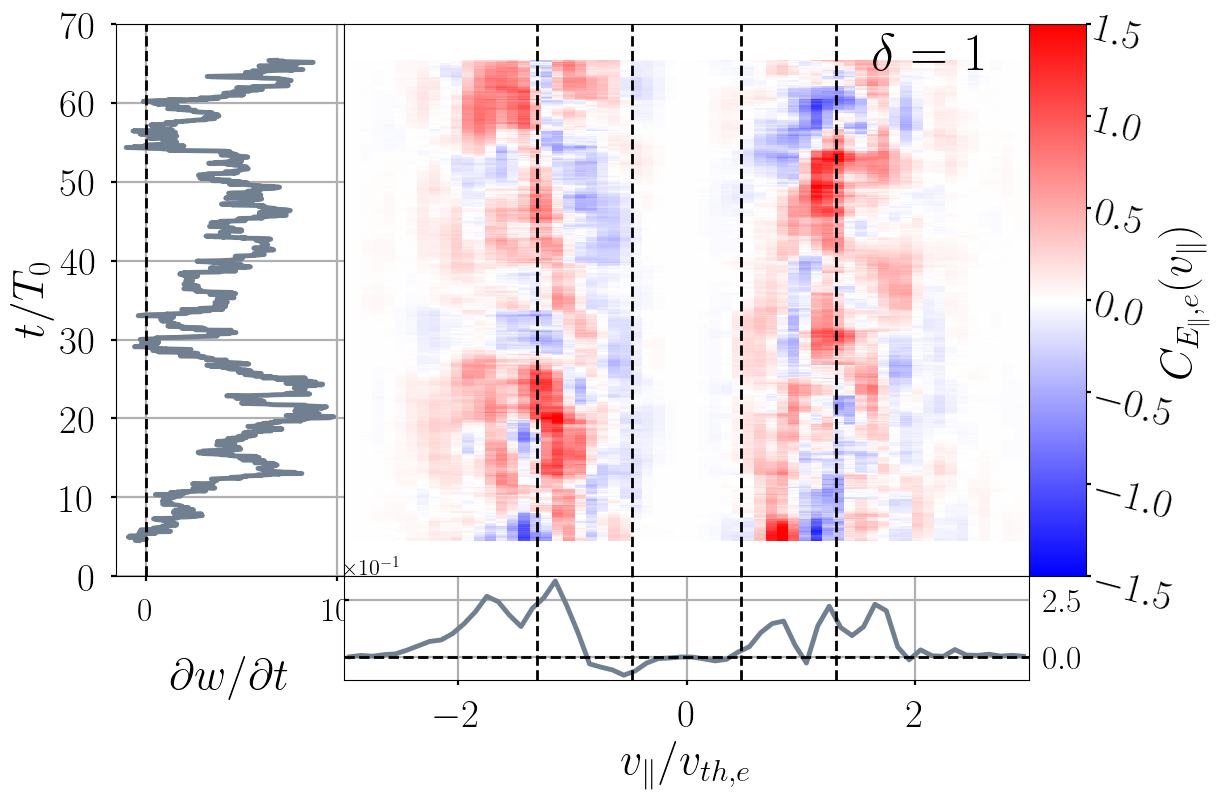}
\includegraphics[width=0.32\textwidth, height=0.25\textwidth]{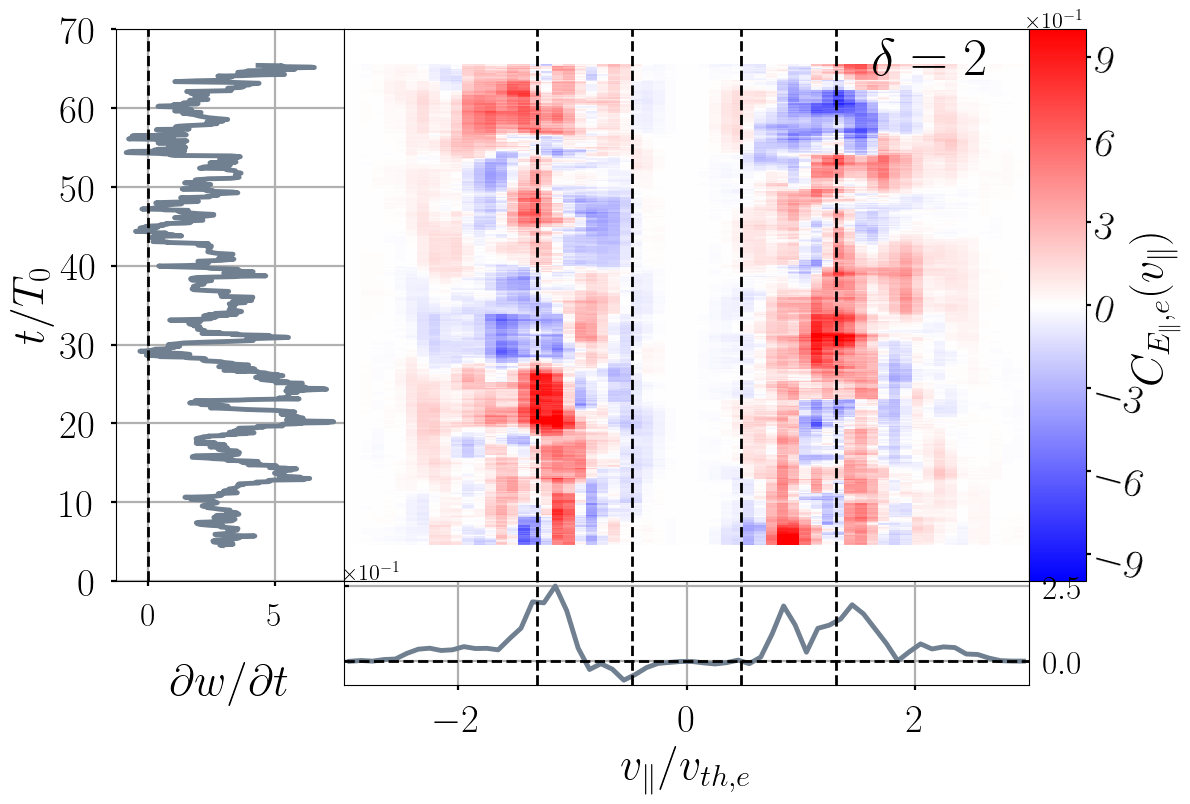}
}
\hfill{
\includegraphics[width=0.32\textwidth, height=0.25\textwidth]{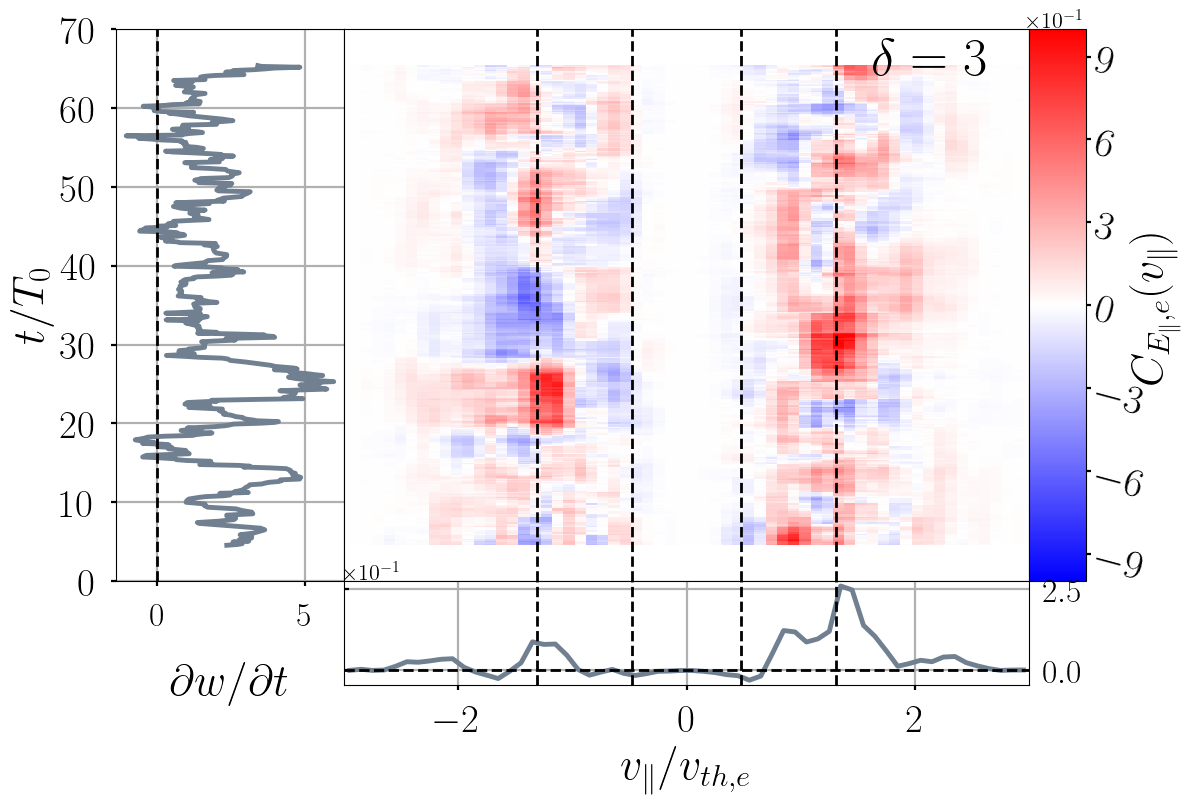}
\includegraphics[width=0.32\textwidth, height=0.25\textwidth]{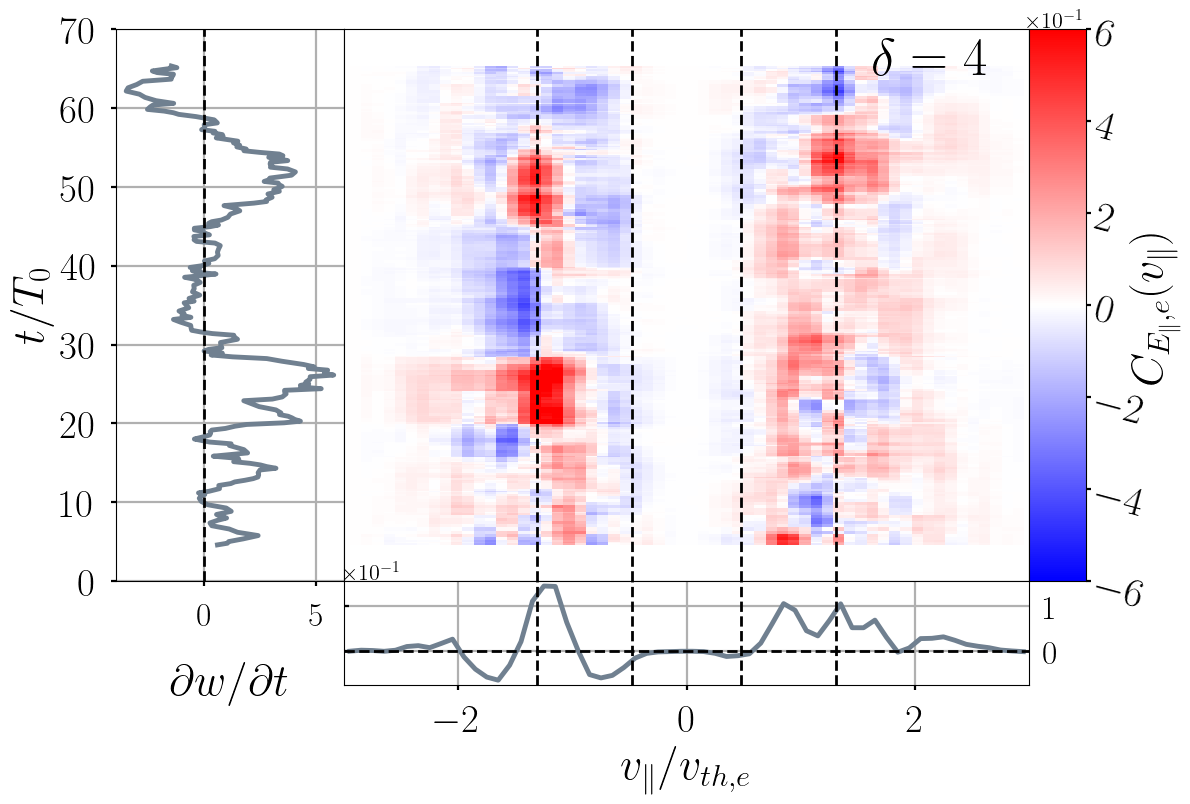}
\includegraphics[width=0.32\textwidth, height=0.25\textwidth]{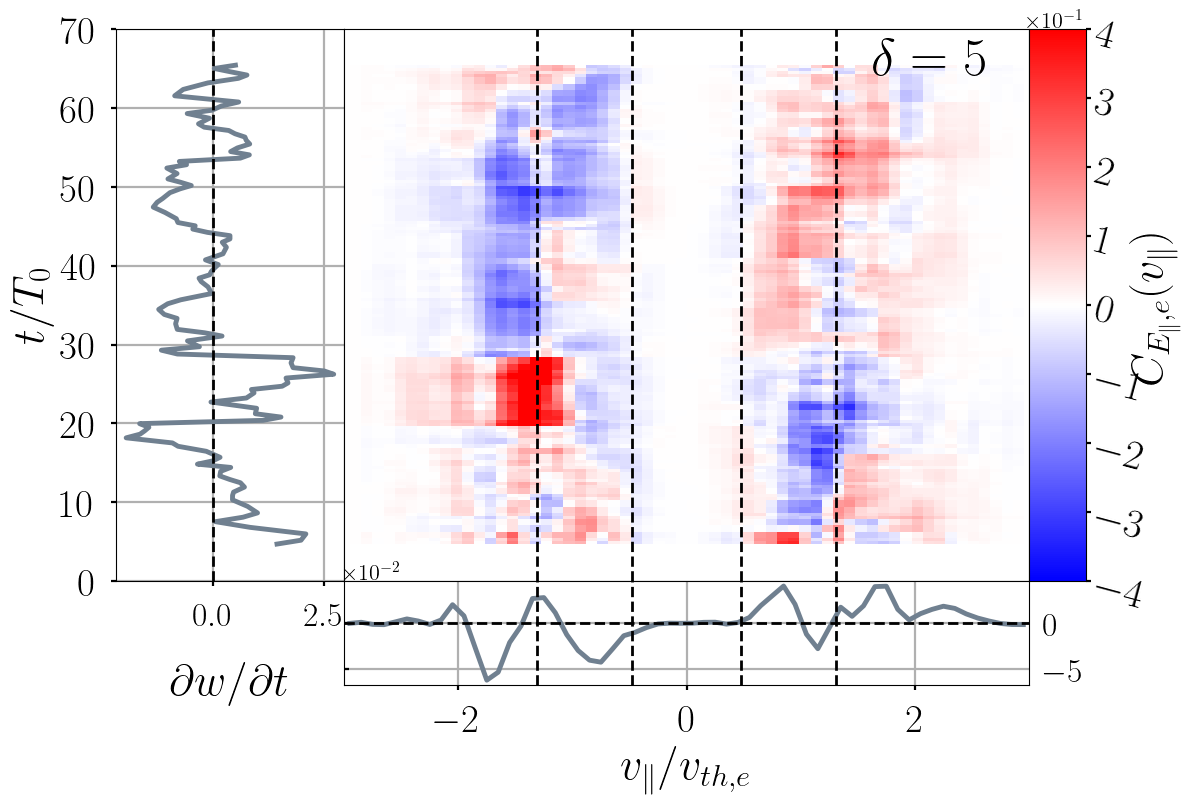}
}
\hfill{
\includegraphics[width=0.32\textwidth, height=0.25\textwidth]{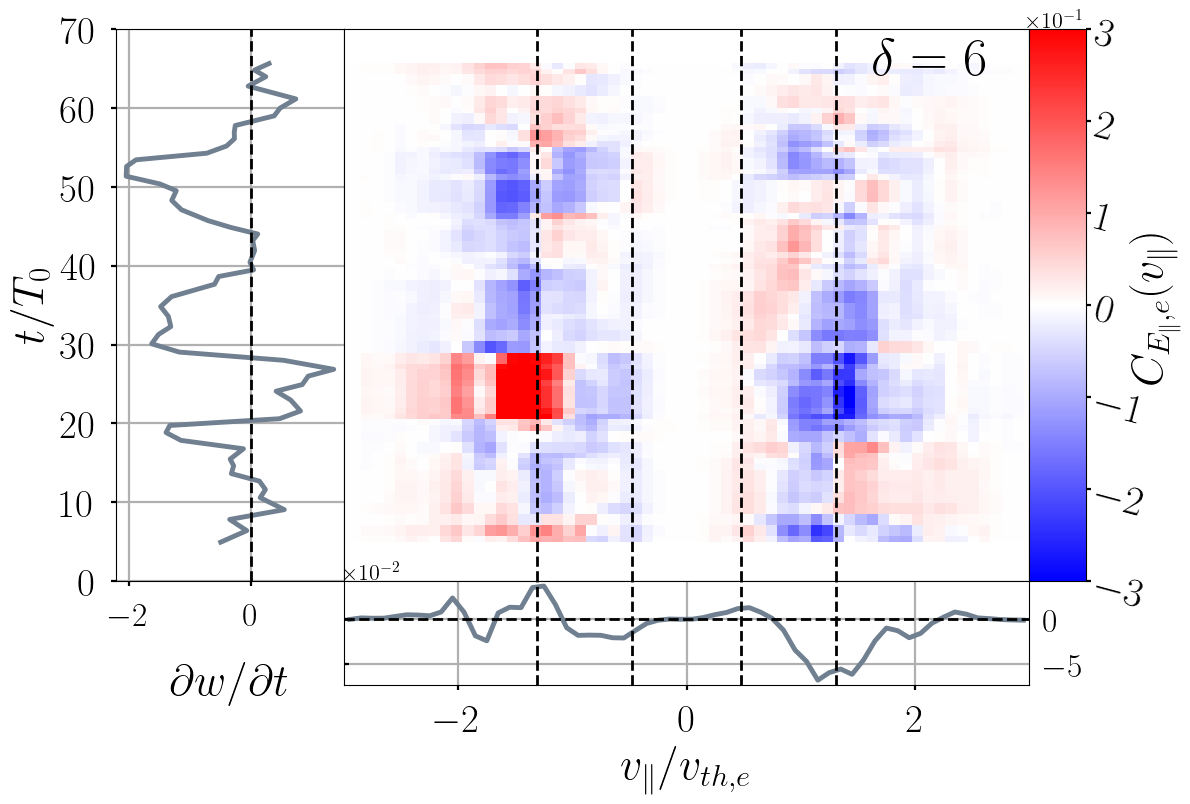}
\includegraphics[width=0.32\textwidth, height=0.25\textwidth]{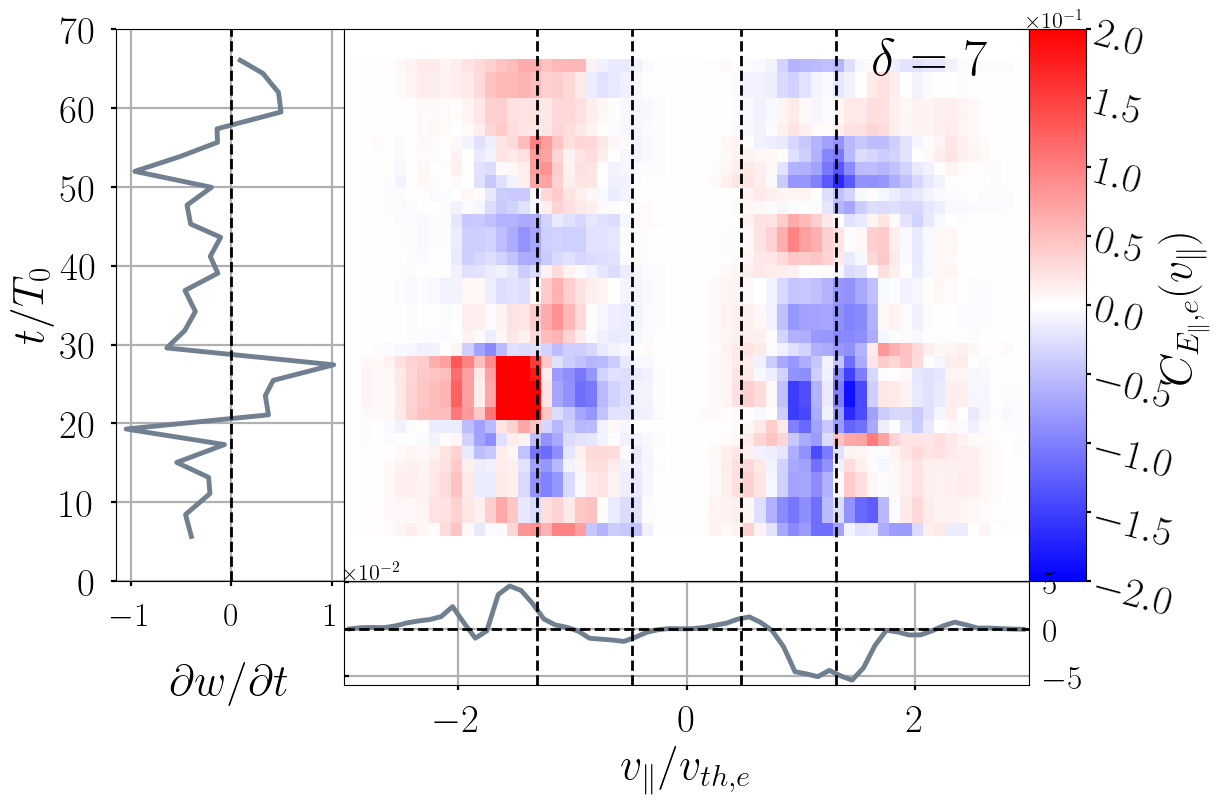}
\includegraphics[width=0.32\textwidth, height=0.25\textwidth]{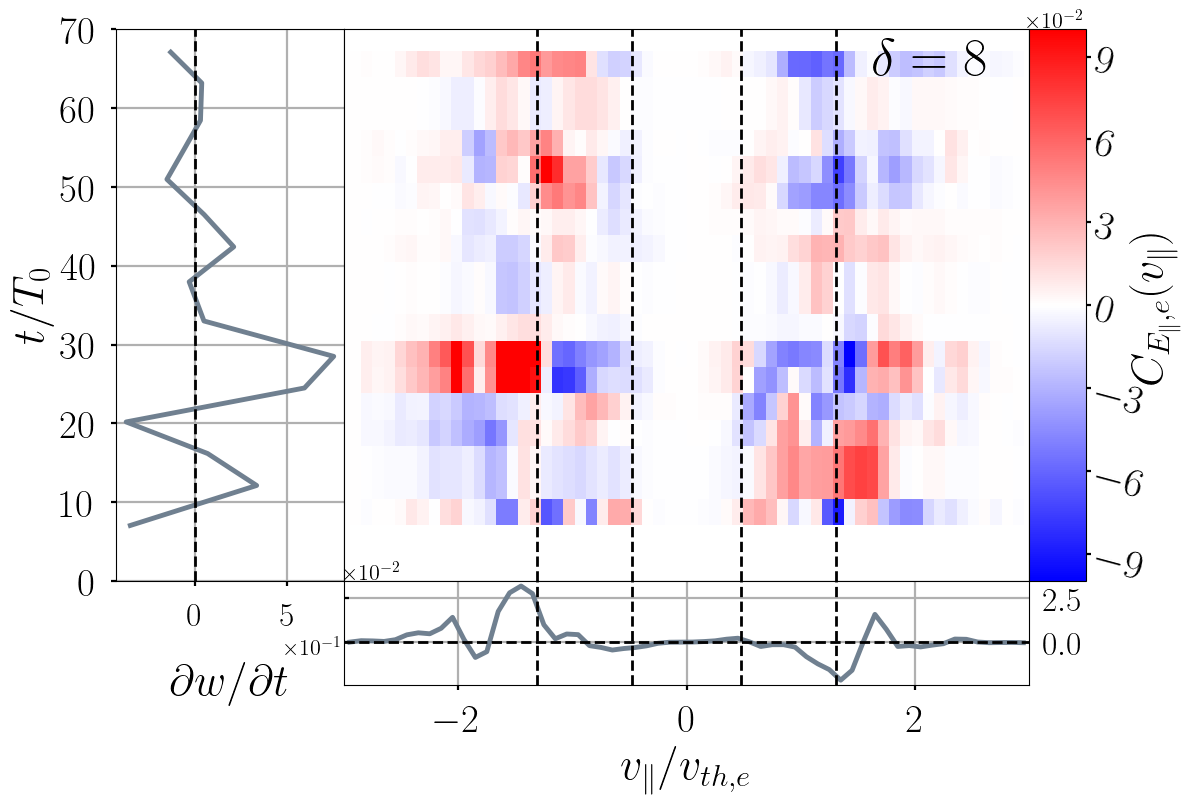}
}
\hfill{
\includegraphics[width=0.32\textwidth, height=0.25\textwidth]{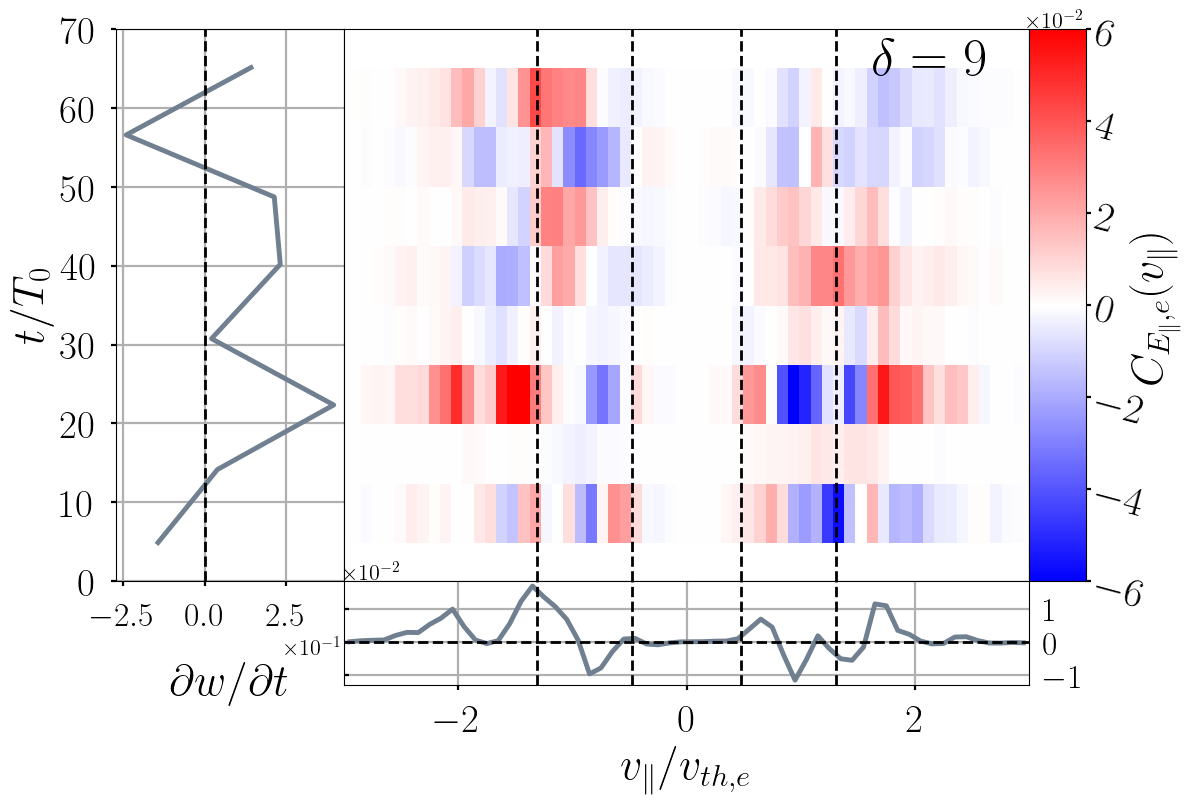}
\includegraphics[width=0.32\textwidth, height=0.25\textwidth]{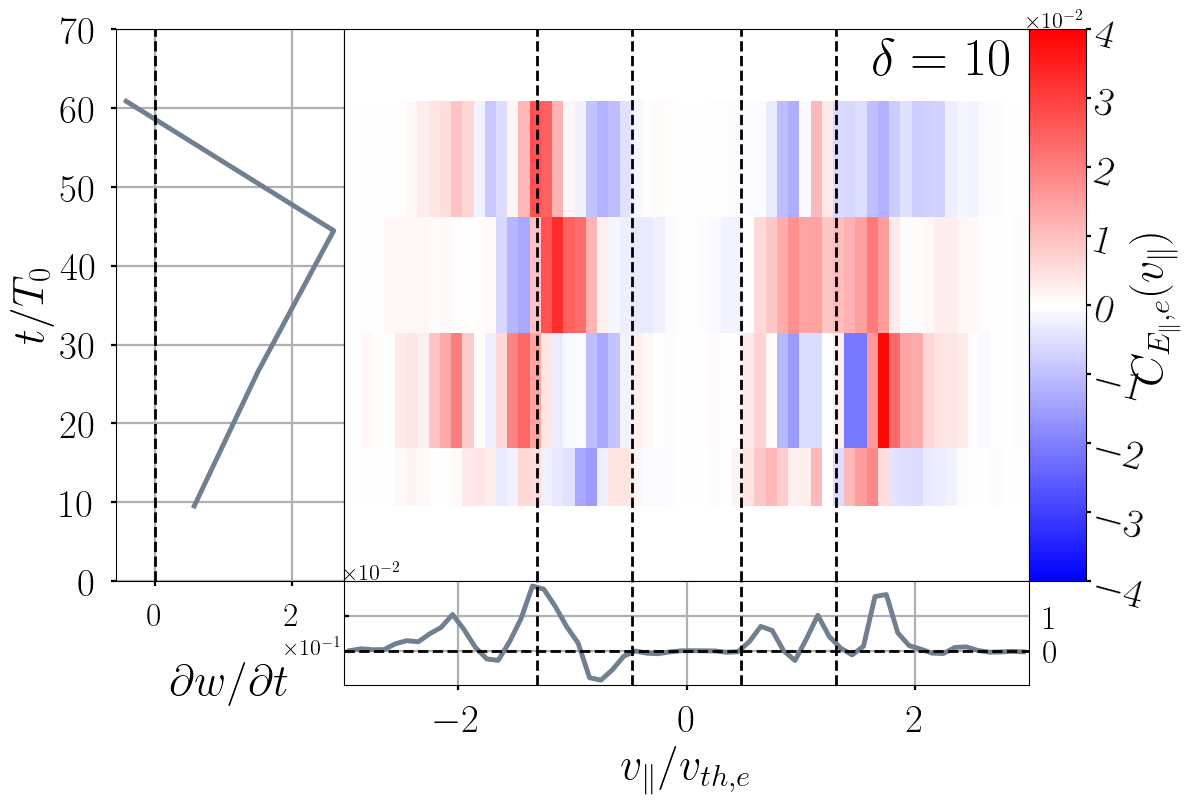}
\includegraphics[width=0.32\textwidth, height=0.25\textwidth]{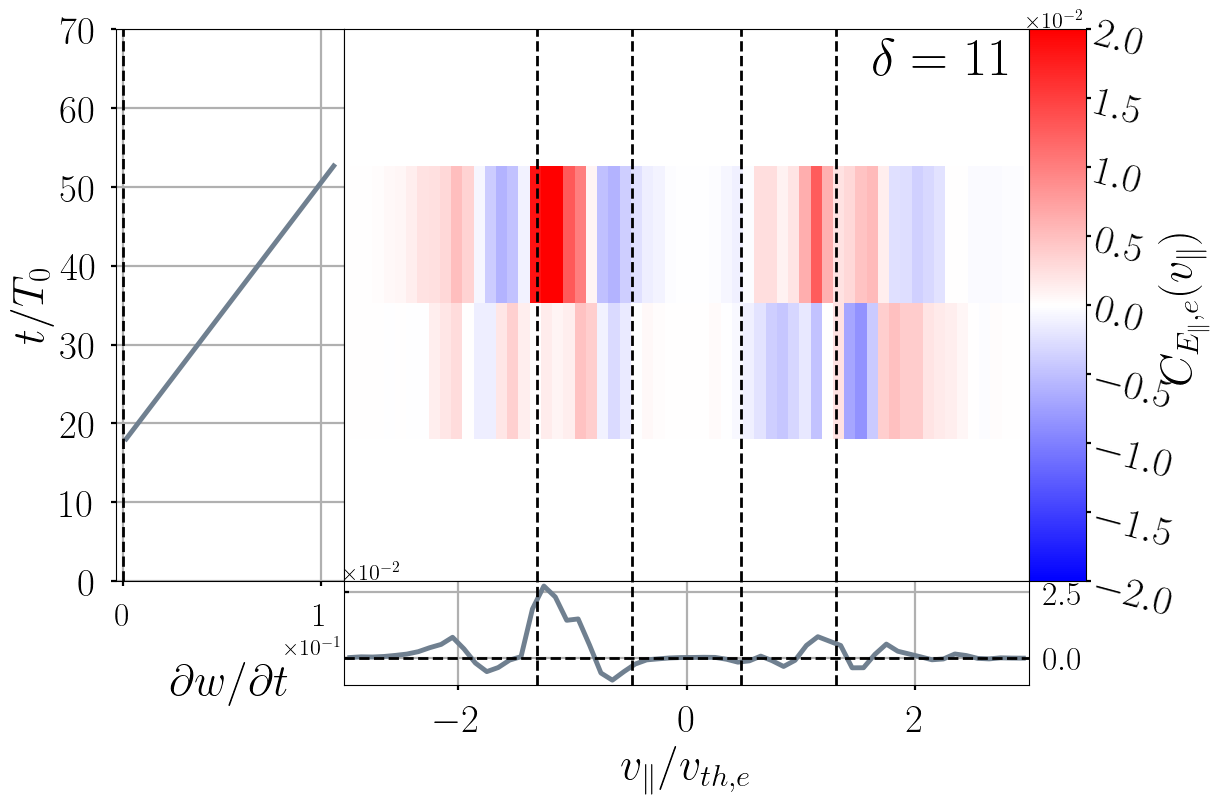}
}
\caption{\label{fig:p8grid} Comparison of the field-particle correlation at simulation probe 8 for datasets $\delta = 0 - 11$, illustrating how a signature changes through decreasing time-resolution. The vertical dashed lines indicate the minimum and maximum parallel phase velocities of the kinetic \Alfven wave over the resolved range of perpendicular scales, as shown in Fig.~\ref{fig:GKDR}.} 
\end{centering}
\end{figure*}

For each downsampling exponent $\delta$ in Table~\ref{tab:res} and at all $24$ probe points in the simulation, we apply the field-particle correlation technique \eqref{eq:cepar} to yield the parallel electric correlation $C_{E_\parallel}(v_\parallel,  v_\perp, t)$: a measure of the rate of change of phase-space energy density as a function of gyrotropic velocity space $(v_\parallel,  v_\perp)$. Bipolar signatures of Landau damping appear in the correlation organized around a value of $v_\parallel$ corresponding to the approximate parallel phase velocity of the damped \Alfven wave.\citep{Klein:2017,Chen:2019,Afshari:2021} These signatures contain little to no perpendicular structure---other than a monotonic drop off of the amplitude as $\exp(-v_\perp^2/v_{te}^2)$---so the gyrotropic correlation may be integrated over $v_\perp$ for each output time, resulting in one-dimensional reduced parallel correlations, $C_{E_\parallel}(v_\parallel, t)$. These reduced parallel correlations may be stitched together in time to create a \emph{timestack plot} that shows the rate of change of the phase-space energy density as the simulation progresses. 

Examples of timestack plots of the reduced parallel correlation $C_{E_\parallel}(v_\parallel, t)$ taken at probe 8 of the simulation are shown in Fig.~\ref{fig:p8grid} (central panels), where a correlation interval of $\tau \approx 8.7\ T_0$ was used. These timestack plots are then averaged over the full time interval to produce 1D correlations as a function of $v_\parallel$ only (lower panels of Fig.~\ref{fig:p8grid}), or over the remaining parallel velocity coordinate to give the rate of change of the energy density due to the work done by the parallel component of the electric field, $\left(\partial W/ \partial t\right)_{E_\parallel}$ (LH panels of Fig.~\ref{fig:p8grid}). All of this is consistent with typical use of the FPC technique.

Taken as a whole, Fig.~\ref{fig:p8grid} illustrates how FPC signatures of Landau damping can be traced through the downsampling process. The correlations of the different datasets are labeled according to their corresponding downsampling exponent, $\delta$, which increases from upper left to lower right. Each parallel velocity axis is normalized to the electron thermal velocity ($v_\parallel/v_{te}$), and each time axis to the period of the domain-scale \Alfven wave ($t/T_0$). The correlation interval in the timestack plots is $\tau/T_0 \approx 8.7$, while in the lower panels showing the one-dimensional reduced parallel correlations it is equal to the full duration of the simulation, $\tau/T_0 = 68.9$.  

When looking to recover FPC signatures from undersampled waves, as we do here in datasets $\delta \geq 4$, the correlation interval $\tau$ is chosen differently than usual. Typically, a length of a few outer-scale wave periods is sufficient to achieve cancellation of oscillatory energy transfer, even for broadband turbulence;\citep{Klein:2016,Klein:2017} here, however, to anticipate the of analysis of the lowest time resolution versions of the data, we choose to focus on the correlations where $\tau$ is equal to the full time interval. As illustrated in Fig.~\ref{fig:f_fNy}, a longer averaging interval will theoretically include a more representative set of all phases of an undersampled frequency. Thus, maximizing $\tau$ should maximize the cancellation of oscillatory energy transfer from the correlation. Therefore, in the analysis below, we focus primarily on the one-dimensional reduced parallel correlations $C_{E_\parallel}(v_\parallel)$, as shown in lower panels of Fig.~\ref{fig:p8grid}.

\begin{figure*}[!t]
\includegraphics[width=0.95\textwidth, height=0.4\textwidth]{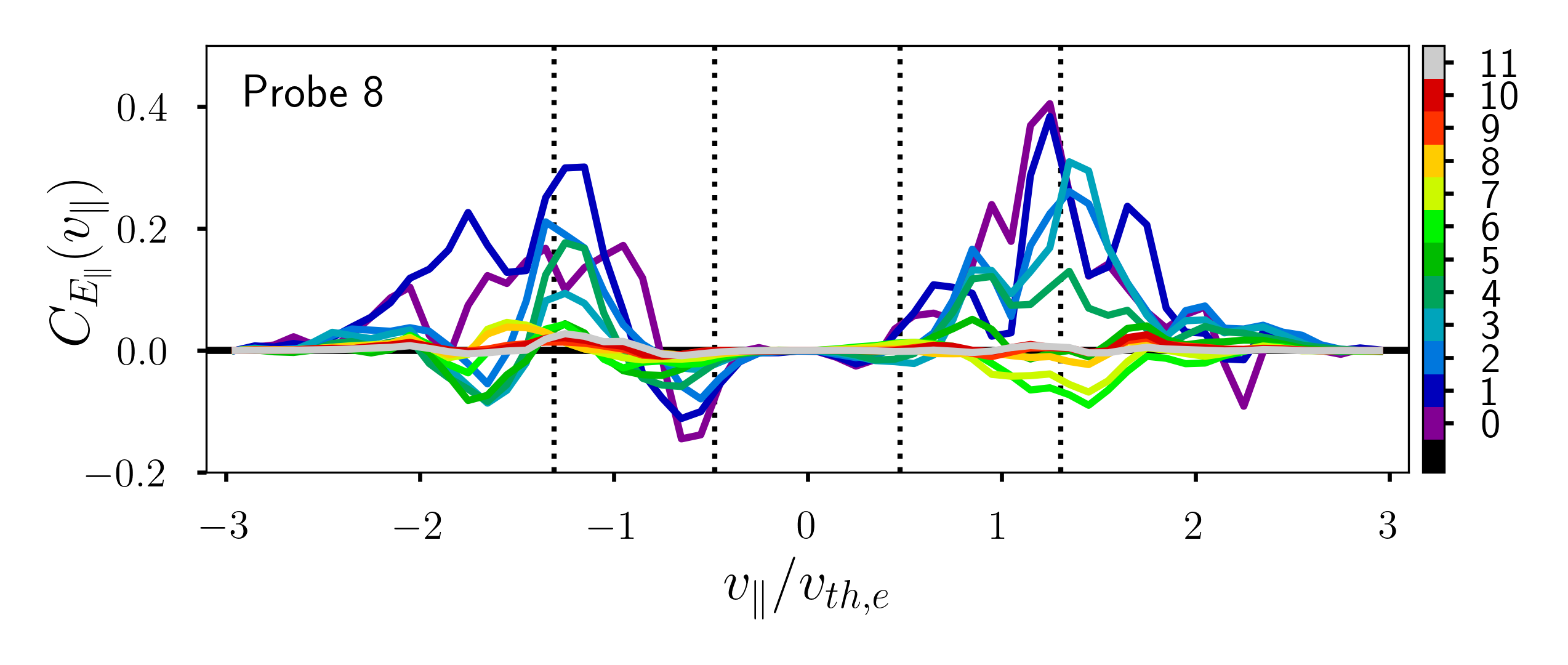}
\caption{\label{fig:intCdt_overlay} Overlay of the time-averaged reduced parallel correlations at probe 8, colored according to downsampling exponent $\delta$. The bipolar signature located at $v_\parallel > 0$ is last visible for $\delta=4$, and the signature for which $v_\parallel < 0$ is retained through $\delta=11$} 
\end{figure*}

A bipolar signature of Landau damping appears in the correlation as a zero-crossing from negative to positive as $|v_\parallel|$ increases, and this zero-crossing occurs at approximately the parallel phase velocity $v_{ph}$ of the damped wave. The bipolar structure arises when particles with parallel velocities less than the phase velocity ($|v_\parallel| < |v_{ph}|$) are accelerated to velocities above the phase velocity ($|v_\parallel| > |v_{ph}|$), leading to a loss of phase-space energy density below $v_{ph}$ (blue) and a gain of phase-space energy density above $v_{ph}$ (red). The signature therefore corresponds to a flattening of the VDF at the resonant velocity, as expected from the quasilinear evolution of Landau damping. \citep{Klein:2017,Howes:2017}

Since \Alfven waves may travel both up and down the background magnetic field, the bipolar signatures may be symmetric about $v_\parallel = 0$, as was seen in the first signature of Landau damping observed \textit{in situ}.\citep{Chen:2019} Due to the transient nature of plasma turbulence and the potential imbalance of the damping of upward and downward propagating waves, however, these signatures are not necessarily symmetric. \citep{Afshari:2021} Additionally, in the case of dispersive waves in broadband turbulence, such as that considered here, the signatures often appear broadened across $v_\parallel$ as a result of the overlap of signatures of Landau damping of kinetic \Alfven waves with different $k_\perp \rho_i$, and therefore different parallel phase velocities\citep{Horvath:2020} (see Fig.~\ref{fig:GKDR}). 

As the downsampling exponent increases, a signature that was initially visible at high time resolution may become lost. Whether some signatures persist at $f/f_{Ny} > 1$, when the damped waves are undersampled, is of primary interest. At probe 8 in dataset $\delta=0$ (indicating a downsampling factor of $2^0 = 1$, or no downsampling) an FPC signature with $v_\parallel < 0$ is visible, highlighted in green in the upper left of Fig.~\ref{fig:p8grid}. The time-integrated reduced parallel correlations from all twelve dataset versions show that this signature persists through $\delta = 11$ (downsampling factor of $2^{11} = 2048$). For a signature to remain visible in the final dataset was rare, occurring in only two cases. Additionally, note that the amplitude of the probe 8, $\delta=11$ signature is dramatically reduced from its original value in $\delta=0$. Considering the low number of samples ($N_{FPC} = 2$) in the lowest resolution dataset, the validity of this appearance of the signature may rightly be questioned. This consideration will be addressed in a later section; for now, we take at face value the appearance of a bipolar signature in any dataset version. 

A possible signature also exists in $v_\parallel > 0$ at probe 8. A clear bipolar pattern is not visible in the time-integrated correlation of the original dataset; however, the color plot of $C_{E_\parallel}(v_\parallel, t)$ indicates that signatures due to Landau damping may occur approximately in the time range between $15 \le t/T_0 \le 25$, and later between $42 \le t/T_0 \le 45$. 
The velocity-space structure during other time intervals in the simulation is potentially obscuring these signatures in the time-integrated plot up until $\delta = 3$, when a zero-crossing is visible. This is an example where an overlap of bipolar signatures at different parallel phase velocities can broaden, or even completely obscure, a given resonant damping signature. If we assume that this is indeed a signature of Landau damping, we observe that it persists until $\delta = 5$, at which point the structure of the signature begins to change markedly.

The persistence of a signature can be concisely evaluated by plotting the lower subplots of Fig.~\ref{fig:p8grid} on a single set of axes as shown in Fig.~\ref{fig:intCdt_overlay}, where each time-integrated correlation is colored according to its downsampling exponent. Again, note that the signature due to the \Alfven wave propagating up the magnetic field ($v_\parallel > 0$) is lost between datasets $\delta=4$ and $\delta=5$, and the signature due to the wave propagating down the magnetic field ($v_\parallel < 0$) remains visible through all versions of the data, though it is reduced significantly in amplitude through the downsampling process. 

The process of analyzing the persistence of field-particle correlation signatures as the time resolution of the data is decreased is repeated for all twenty-four probe points. We describe in detail our partly-automated procedure for determining the dataset in which a bipolar signature is lost in Appendix~\ref{app:auto}.

\section{ Results }
\label{sec:results}

\begin{figure}[b!]
\includegraphics[width=0.5\textwidth, height=0.45\textwidth]{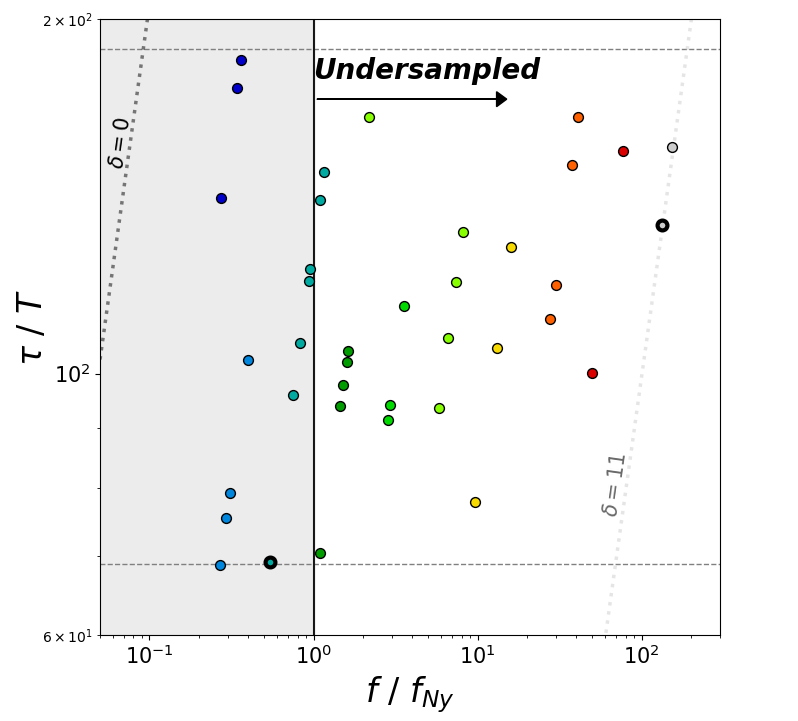}
\caption{\label{fig:taufreq_ao} For each energization signature, the resonant \Alfven wave frequency is normalized by the Nyquist frequency of the dataset in which the signature last appeared ($f/f_{Ny}$) and plotted against the number of wave periods spanned by the correlation interval ($\tau/T$).  }
\end{figure}

The first goal of this work is to determine if particle energization (specifically via Landau damping) can be detected \textit{in situ} using field-particle correlations when the waves that are resonantly damped have frequencies above the Nyquist frequency of the spacecraft instrumentation, $f>f_{Ny}$. As described above, we approach this question by analyzing the persistence of velocity-space signatures of energization in a simulated dataset as the time resolution is systematically decreased. Our method, described in Appendix~\ref{app:auto}, results in a set of $38$ data points located in $(f/f_{Ny}, \tau/T)$ parameter space, where $f = 1/T$ is the estimated frequency of the damped wave, $f_{Ny}$ is the Nyquist frequency of the most extremely downsampled dataset in which a given velocity-space signature remains visible, and $\tau$ is the correlation interval. These parameters are used as dimensionless coordinate axes in Fig.~\ref{fig:taufreq_ao}, which compares the persistence of all $38$ signatures. The data points are color coded by the downsampling factor of the final dataset in which the energization signature appears, so that the value of $f_{Ny}$ is constant for a given color.

The vertical axis in Fig.~\ref{fig:taufreq_ao} denotes the number of wave periods $T$---determined from the estimated frequency $f$ of the damped wave generating an FPC signature---that are spanned by the correlation interval $\tau$. Thus, for a fixed $\tau$, the vertical extent of this plot is limited by the finite range of frequencies that are likely to experience damping. The kinetic \Alfven waves at the domain scale, having the lowest frequency ($\omega_{0}/k_\parallel v_{te}=0.477$) and maximum period ($T=T_0$) in the simulation, fit into the interval $68.9$ times, and so will appear at $\tau/T_0 = 68.9$ (lower dashed line). The highest frequency kinetic \Alfven waves in the simulation ($\omega_{max}/k_\parallel v_{te}=1.31$) correspond to a value of $\tau/T_{max} = (\tau/T_0) (T_0/T_{max}) = 189$ (upper dashed line). Here we have assumed no change in the parallel wavenumber, $k_\parallel= k_{\parallel 0}$, which leads to a lower limit on the frequency of the damped wave $f$, and thus a conservative estimate of the super-Nyquist factor, $f/f_{Ny}$.

The horizontal axis, $f/f_{Ny}$, is the ratio of the estimated frequency $f$ of the KAW undergoing collisionless damping to the Nyquist frequency of the sampling $f_{Ny}$.  The frequency ratio is equivalent to $f/f_{Ny}=2/n$, where $n$ is the number of samples per period of the collisionlessly damped wave. Thus, the horizontal location of a data point directly corresponds to the persistence of an energization signature through the downsampling process. The vertical line at $f/f_{Ny} = 1$ separates the region of well-sampled frequencies to the left (shaded) from the region of undersampled frequencies to the right. Typically, for discretely  sampled data, only those frequencies in the shaded region are accessible. 

The diagonal dotted lines indicate the relation between the number of wave periods in the correlation interval $\tau/T$ as a function of the wave frequency relative to the Nyquist frequency $f/f_{Ny}$ for a fixed sampling cadence over the minimum to maximum wave frequency range, $f_0 \le f \le f_{max}$.  Note that this relation can be expressed by 
\begin{equation}
\frac{\tau}{T} = \frac{1}{2} \left(\frac{\tau}{T_0}\right)\left(\frac{T_0}{\Delta t}\right) \frac{f}{f_{Ny}},
\label{eq:yvsx}
\end{equation}
where $\tau/T_0=68.9$ in the length of the correlation interval in minimum wave periods and $\Delta t/T_0$ is the sampling cadence given in Table~\ref{tab:res}.  Note that the sampling cadence increases by factors of two given by the downsampling exponent $\delta$, such that $\Delta t = \Delta t_0 2^\delta$, where the minimum cadence is $\Delta t_0/T_0=0.017$.  Substituting this relation for $\Delta t$ into \eqref{eq:yvsx} and taking the logarithm gives the equation for the linear relation on a log-log plot as
\begin{equation}
\log \left(\frac{\tau}{T}\right) = \log \left(\frac{f}{f_{Ny}} \right)+
\log \left( \frac{1}{2} \frac{\tau}{T_0}\frac{T_0}{\Delta t_0}\right) - \delta \log 2
\label{eq:logline}
\end{equation}
Therefore, the diagonal dotted lines have a slope of 1 and a $y$-intercept that changes with the downsampling factor $\delta$. The left-hand dotted line corresponds to $\delta=0$, or no downsampling, and the right-hand dotted line corresponds to maximum downsampling with $\delta=11$. Note that no data points appear on the lines representing $\tau/T$ vs. $f/f_{Ny}$ for $\delta = 0$ (or $\delta =1$) since all $38$ of the signatures persist through more than two downsampling passes.  Two points appear on the line for $\delta =11$, corresponding to the two signatures that persist through the final dataset version. 

\begin{figure*}[ht!]
\includegraphics[width=0.3\textwidth, height=0.3\textwidth]{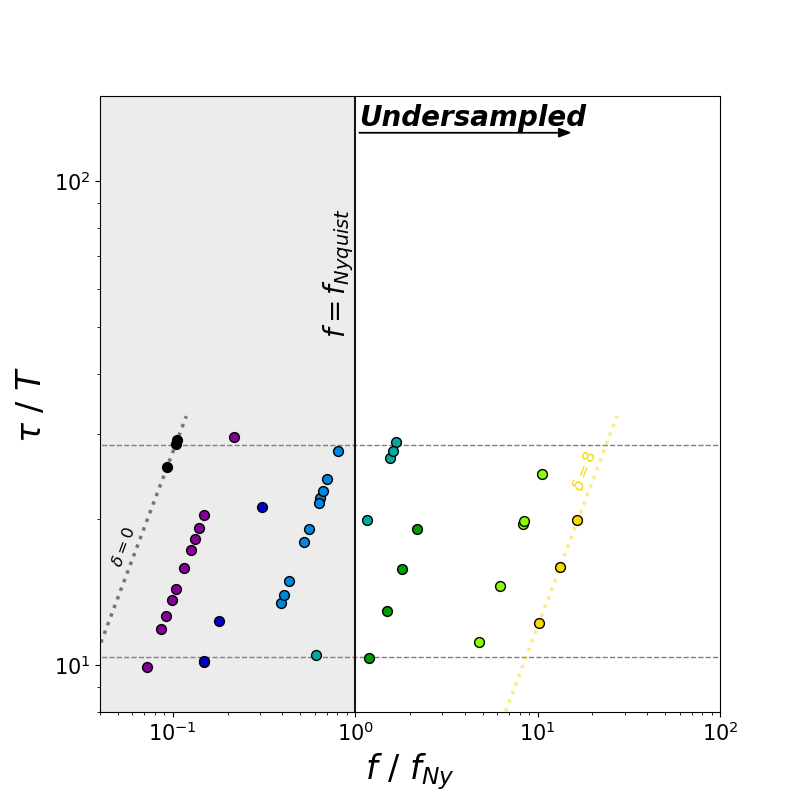}
\includegraphics[width=0.3\textwidth, height=0.3\textwidth]{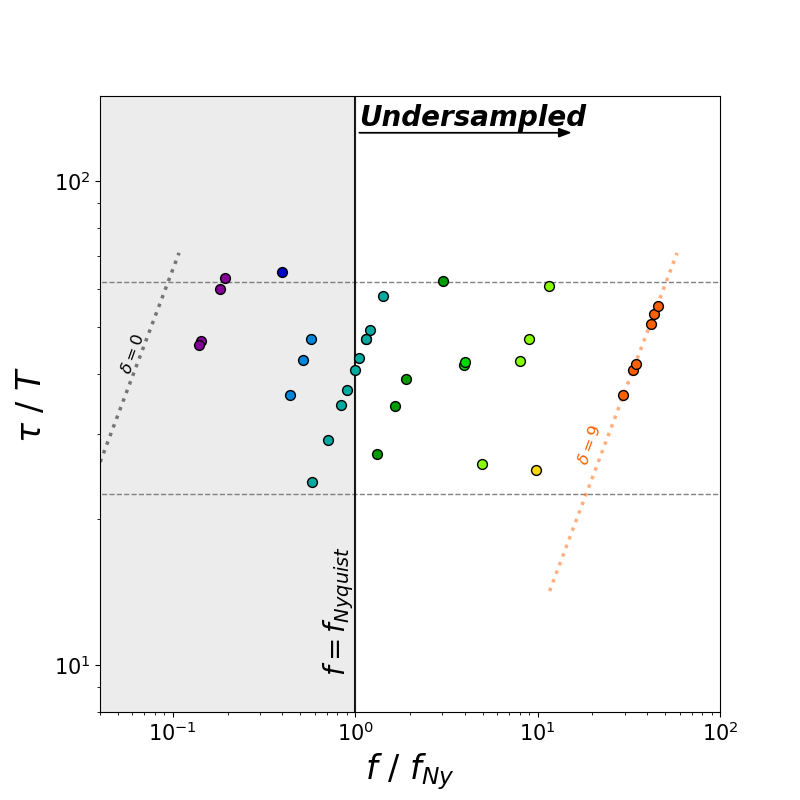}
\includegraphics[width=0.3\textwidth, height=0.3\textwidth]{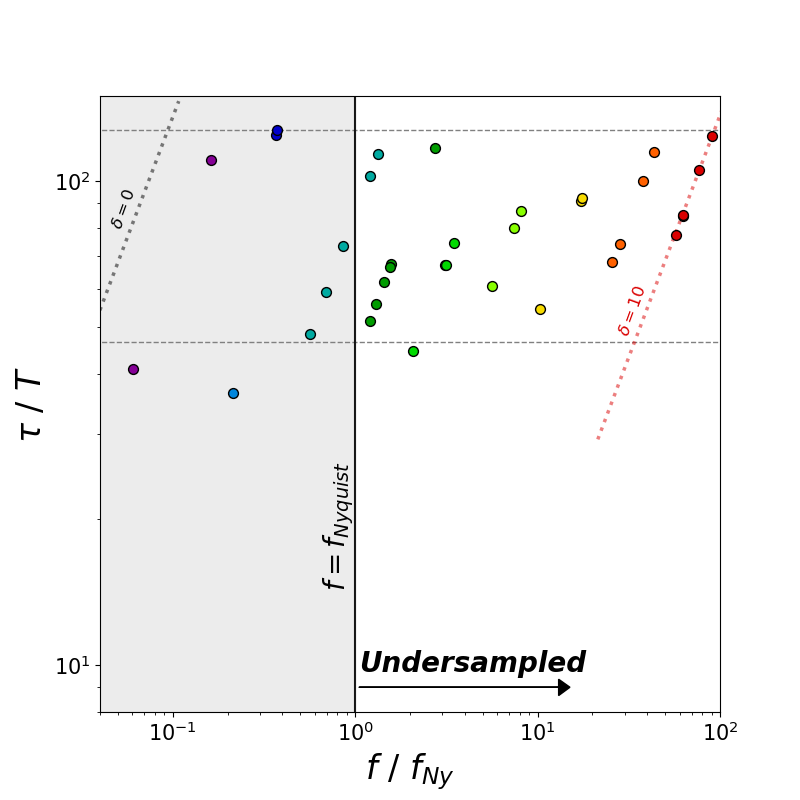}
\caption{\label{fig:taufreq_comp} For each plot the data points are generated using the same method as for Fig.~\ref{fig:taufreq_ao}, but the total length of the simulated time series has been decreased, thereby decreasing $\tau$. From left to right, the length of $\tau$ in terms of the largest wave period is: $\tau/T_0 = 10.4, 22.6, 46.6$.}
\end{figure*}

To illustrate how the last discernible energization signatures are plotted on the $(f/f_{Ny},\tau/T)$ parameter space, the points representing the two FPC signatures at probe 8 are marked with bold outlines in Fig.~\ref{fig:taufreq_ao}. Recall from Figs.~\ref{fig:p8grid} and~\ref{fig:intCdt_overlay} that the signature with $v_{ph} > 0$ was last visible in version $\delta = 4$.  The zero crossing of that signature appears at approximately the minimum KAW parallel phase velocity $\omega/k_\parallel v_{te} \simeq 0.477$.  One may compute the coordinates in $(f/f_{Ny},\tau/T)$ parameter space from this observed zero-crossing phase velocity by 
\begin{equation}
    \frac{f}{f_{Ny}}=2 \left(\frac{\Delta t}{T_0}\right)
    \left( \frac{\omega}{k_\parallel v_{te}}\right)
   \left( \frac{k_\parallel v_{te}}{\omega_0}\right)
   \label{eq:ffNy}
\end{equation}
\begin{equation}
    \frac{\tau}{T}= \left(\frac{\tau}{T_0}\right)
    \left( \frac{\omega}{k_\parallel v_{te}}\right)
   \left( \frac{k_\parallel v_{te}}{\omega_0}\right)
   \label{eq:tauT}
\end{equation}
where  $\omega_0/k_\parallel v_{te}= 0.477$, $ \tau/T_0=68.9$,
and the normalized sampling cadence for $\delta=4$ is given in Table~\ref{tab:res} by $\Delta t/T_0=0.268$.  The resulting point in parameter space is $(f/f_{Ny},\tau/T)=(0.536,68.9)$, as shown by the lower point with bold outline, indicating that the signature is lost while well sampled and consequently appears in the shaded region to the left of the $f=f_{Ny}$ threshold.  Note that, for downsampling exponent  $\delta=4$, the limits of $f/f_{Ny}$ are given in 
Table~\ref{tab:res} by $f_0/f_{Ny} = 0.534$ and $f_{max}/f_{Ny} = 1.48$.  For the probe 8 signature with $v_{ph} < 0$, recall that it persists through the final downsampled version of the data, with a zero crossing at approximately $\omega/k_\parallel v_{te} \simeq 0.93$.   This yields a position in parameter space of $(f/f_{Ny},\tau/T)=(134,134)$, indicating that the energization signature is recovered for the damping of a wave with a frequency more than 100 times greater than the Nyquist frequency, 
 $f/f_{Ny} > 100$. Note that  Table~\ref{tab:res} gives the $f/f_{Ny}$ limits for $\delta=11$ of $f_0/f_{Ny} = 34.6$ and $f_{max}/f_{Ny} = 189$. 

Of the $38$ signatures that appeared in the original dataset, $68\%$ remain recoverable by the FPC technique after the data were downsampled to a degree that the Nyquist frequency fell below the frequency of the wave being damped. Those points appear in the unshaded region in Fig.~\ref{fig:taufreq_ao}. This key result---obtaining $26$ data points to the right of $f / f_{Ny}=1$---answers our first science question, demonstrating that the field-particle correlation technique can indeed recover the velocity-space signatures of electron Landau damping of waves that are undersampled relative to the Nyquist frequnency, $f / f_{Ny}>1$.

Finally, it is instructive to consider the effect of adjusting the correlation interval on the persistence of the velocity-space signatures. The simulation length constrains us to a maximum correlation interval $\tau$, but we may create shortened versions of the full time series and perform the same analysis. The results of three such shortened intervals are shown in Fig.~\ref{fig:taufreq_comp}. From left to right these intervals span from $\left(\tau/T_0, \tau/T_{min}\right) = \left(10.4, 28.5\right)$, $\left(22.6, 61.9\right)$, and $\left(46.6, 127.5\right)$. The data points are determined using the same procedure as described above for Fig.~\ref{fig:taufreq_ao}. In the shortest of these three time series, $42\%$ of the signatures persist beyond the Nyquist frequency boundary at $f/f_{Ny} > 1$. In the mid-length interval, $69\%$ of signatures persist into the undersampled region, and $71\%$ in the longest. Taken together, there appears to be a trend for longer correlation intervals to result in higher percentages of signatures that remain visible despite being undersampled. 

\section{Discussion}

As discussed in Sec.~\ref{sec:Qs}, averaging over a greater number of wave periods increases the accuracy that may be achieved in canceling oscillatory energy transfer from the field-particle correlations of undersampled systems. Theoretically this implies the existence of a trend from lower left to upper right in Fig.~\ref{fig:taufreq_ao}: as the frequency and number of wave periods spanned by a fixed $\tau$ increases, an FPC signature should be more resilient to increasing $f/f_{Ny}$. Such an upward trend would supply the rule of thumb that is the second goal of this work. Specifically, given an approximate ratio of $f/f_{Ny} > 1$ for an \textit{in situ} dataset, we would like to predict what length of $\tau$ will be sufficient for recovering FPC signatures. However, though an upward trend is somewhat visible in the data from our simulation, it is not particularly strong nor strictly followed.  Below we provide a empirically derived rule of thumb for the minimum correlation interval needed to obtain a robust FPC signature, discuss some of the physical considerations involved in the collisionless damping of strong plasma turbulence that may impact our results, and apply that rule of thumb to previous and planned observational analyses of spacecraft measurements.

\subsection{Rule of Thumb for Required Interval Length}

The second key question driving this study is to determine a rule of thumb that, given the undersampling factor $f/f_{Ny}>1$ for a wave with an estimated frequency $f$, provides an estimate for the minimum correlation interval $\tau$ needed to reliably recover the velocity-space signature of a particular particle energization mechanism. This rule of thumb can be written in the following simple form
\begin{equation}
    \frac{\tau}{T} \gtrsim \frac{\mathcal{N}}{2} \frac{f}{f_{Ny}},
    \label{eq:rule}
\end{equation}
where we can empirically determine the threshold constant $\mathcal{N}$ for the 
length of the correlation interval in terms of the wave period $T$.
Using the definition for the Nyquist frequency $f_{Ny}=1/(2\Delta t)$, this equation can easily be manipulated into the form 
\begin{equation}
    \frac{\tau}{\Delta t} \gtrsim \mathcal{N},
\end{equation}
where the physical interpretation of threshold constant $\mathcal{N}$ is the total number of samples over the full correlation interval.

In Appendix~\ref{app:thumb}, we describe a mathematical model used to determine empirically the value of the threshold constant needed to recover a reliable energization signature, obtaining a value of $\mathcal{N}=12$.  This rule of thumb yields the minimum length of the correlation interval (in terms of wave periods) $\tau/T$ given an undersampling factor $f/f_{Ny}>1$, but additional physical considerations discussed below imply that this represents a lower bound on the necessary correlation interval. 

\subsection{Physical Considerations}
Whether the rule of thumb given by \eqref{eq:rule} provides useful guidance for determining the minimum correlation length needed to recover a meaningful energization signature requires one to take into account several key physical considerations:
(i) the persistence of kinetic \Alfven wave energy flux throughout the full correlation interval; (ii) the simplifying assumption that there is no parallel cascade of energy; and (iii) the amplitude ratio of the oscillatory to the secular energy transfer associated with a particular energization mechanism.

\begin{figure}[b!]
\includegraphics[width=0.5\textwidth, height=0.35\textwidth]{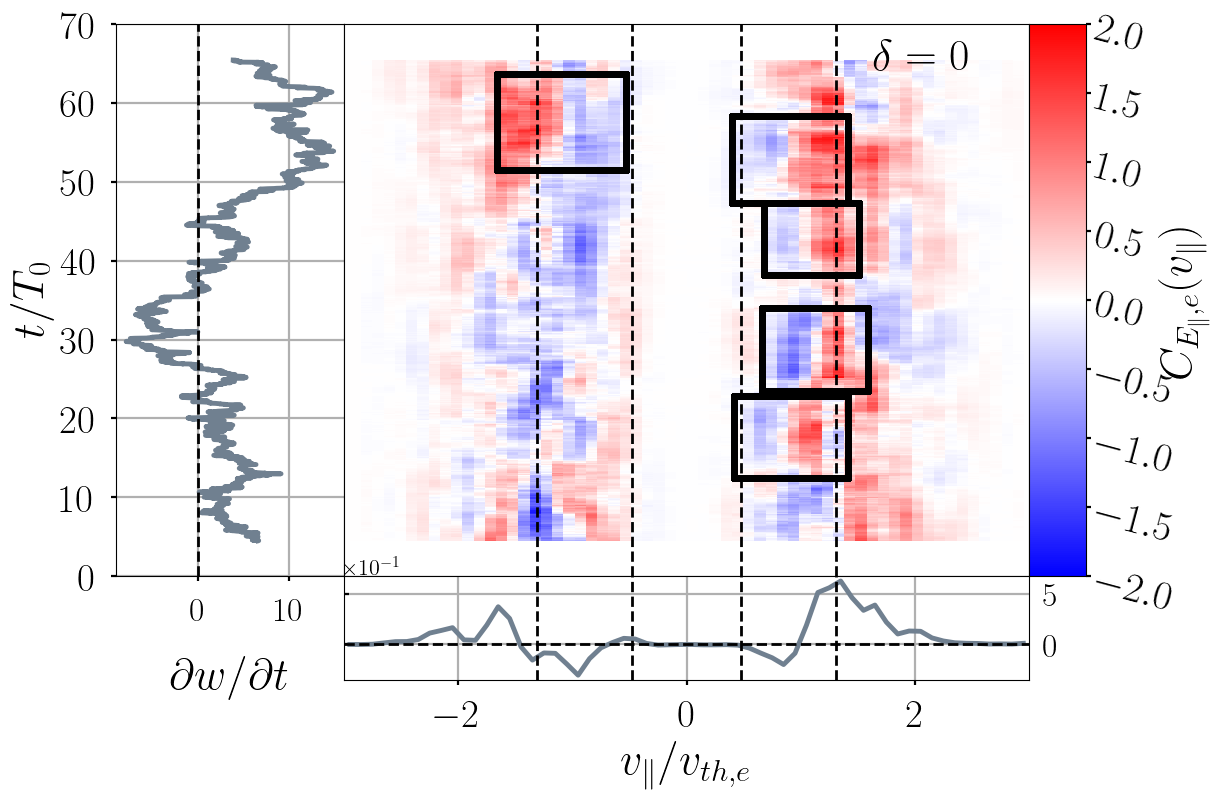}
\caption{\label{fig:p14} Field-particle correlation of dataset $\delta=0$ at probe 14, showing evidence of intermittent Landau damping. }
\end{figure}

Although a steady state turbulent cascade will have a relatively constant level of power in the energy spectrum as a function of frequency or wavenumber \emph{when averaged over a sufficiently long time or large volume}, at a single spatial position the energy in the turbulent fluctuations at a given wavenumber is expected to vary significantly on the timescale of the damping of those fluctuations---such temporal intermittency of the turbulent fluctuations has recently been  explored in MHD and kinetic turbulence simulations.\citep{TenBarge:2013a,Zhdankin:2015a,Zhdankin:2015b}  Since the field-particle correlation technique is computed locally at a single point in space, the collisionless damping and resulting particle energization will have a similar variability in time.  Therefore, it is improbable that the waves causing each of the $38$ signatures in  Fig.~\ref{fig:taufreq_ao} experienced Landau damping over the full simulated interval. 

An example of such temporal intermittency can be seen in the timestack plot of the reduced parallel correlation at probe 14, shown in Fig.~\ref{fig:p14}. The bipolar signature in the lower panel for $v_{\parallel} > 0$ is composed of the superposition of four signatures: each lasting about $8-10\ T_0$ (boxed). Though these are not all at the same phase velocity, they combine into a single bipolar signature in the time-integrated reduced parallel correlation at the bottom of the figure. By contrast, though a bipolar signature appears in $v_{\parallel} < 0$ near the end of the time interval with a similar duration to the four $v_{\parallel}>0$ signatures, no corresponding signature appears in the time-integrated reduced parallel correlation in the lower panel. 

If a kinetic \Alfven wave that is undergoing collisionless damping at a fixed spatial position exists for only a fraction of the total simulated time interval, the effective correlation interval for the associated signature would be only as long as the wave persisted, less than the total correlation interval $\tau$. Thus, the position in $(f/f_{Ny}, \tau/T)$ parameter space marking where that energization signature ceases to be observable as the downsampling increases should rightly be at a lower value of $\tau/T$. This means that the vertical location of the points plotted in Fig.~\ref{fig:taufreq_ao} should be considered as an upper bound. 

For example, in Fig.~\ref{fig:model_phi45} of Appendix~\ref{app:thumb}, the rule of thumb given by \eqref{eq:rule} with  $\mathcal{N}=12$ is plotted as a solid red line. The points from Fig.~\ref{fig:taufreq_ao} overplotted in Fig.~\ref{fig:model_phi45} at $\log(f/f_{Ny})<1$ fall to the left and above this rule of thumb. Ideally,  for a fixed value of $\tau/T$, as the downsampling is increased the point will move to the right with increasing $f/f_{Ny}$.  The position of the plotted point in the $(f/f_{Ny}, \tau/T)$ parameter space, which represents where the energization signature is last detectable, should fall when the downsampling causes $f/f_{Ny}$ to reach the rule of thumb (red line).  An explanation of any point that  appears to the left of this threshold is that temporal intermittency leads to a smaller effective correlation interval: thus, that point should have been at a  lower position in $\tau/T$. 

A second consideration is to understand the effect of neglecting the parallel cascade. In strong MHD turbulence at $k_\perp \rho_i \ll 1$, the wavevector anisotropy is predicted to scale according to $k_\parallel \propto k_\perp^{2/3}$;\citep{Goldreich:1995} in the regime of kinetic \Alfven wave turbulence ($k_\perp \rho_i \gg 1$), the predicted scaling is weaker with $k_\parallel \propto k_\perp^{1/3}$.\citep{Howes:2011b}  As stated earlier in Sec.~\ref{sec:time}, for our simulation in the KAW turbulence regime, neglect of the parallel cascade implies that the individual wave frequencies may be underestimated up to a factor of about two, moving the position of points in $(f/f_{Ny}, \tau/T)$ parameter space.  A factor of two increase in the frequency will move a point to the right by that factor; similarly, the factor of two decrease in the wave period will shift the point up by the same factor. The result is that the point will move along a line of slope $1$ in  $(f/f_{Ny}, \tau/T)$ parameter space, parallel to the rule of thumb given by \eqref{eq:rule}. Therefore, neglect of the parallel cascade does \emph{not} impact the position of the point relative to the rule of thumb threshold.
 
A third physical consideration is that the ability of an undersampled signal to recover a particle energization signature depends on the relative amplitude of the secular energy transfer (associated with collisionless damping) to the oscillating energy transfer (associated with undamped wave motion).\citep{Klein:2016,Howes:2017}  For a smaller ratio of the two quantities, it stands to reason that one may need a longer correlation interval to achieve a more complete cancellation of the oscillating energy transfer, thus exposing the smaller amplitude secular energy transfer. As shown in Fig.~\ref{fig:GKDR}, the normalized rate of electron Landau damping $-\gamma/\omega$, which quantifies the relative strength of the collisionless damping, increases monotonically over the range of scales in the simulation. Thus, at sufficiently small scales, the bipolar velocity-space signature of  electron Landau damping will have an increasing contribution relative to the oscillating energy transfer, and thus the required correlation interval relative to the wave period may decrease. The details of this balance of collisionless damping to the oscillating energy transfer associated with undamped wave motion depend on the nonlinear kinetic plasma physics of the turbulence, and so numerical simulations are essential to assess the length of correlation interval needed for recovering signatures of a particular physical mechanism. Other proposed turbulent dissipation mechanisms may require a longer or shorter threshold correlation interval. Here we take the specific case of electron Landau damping to propose an initial rule of thumb for the recovery of energization signatures from undersampled plasma turbulence.

A final point is that the rule of thumb provides a new means of assessing the statistical significance of any energization signature generated by field-particle correlation analysis.  Any points in  $(f/f_{Ny}, \tau/T)$ parameter space that fall to the right of and below the rule of thumb, such as points for downsampling exponents $\delta=10$ and $\delta=11$  in Fig.~\ref{fig:model_phi45}, should be treated with caution as they may not be statistically significant. The empirically determined threshold value of $\mathcal{N}=12$ in \eqref{eq:rule} suggests one needs at least 12 samples to obtain a reliable velocity-space signature; since Table~\ref{tab:res} shows that downsampling exponents $\delta=10$ and $\delta=11$ have 4 and 2 samples, respectively, signatures at such extreme downsampling exponents may not be physically meaningful.  In such cases, one may use the procedure of phase-randomization of the fields, as described in Chen \emph{et al.} (2019),\citep{Chen:2019} to assess whether a given velocity-space signature is statistically significant.

\subsection{Application to Undersampled \textit{in situ} Spacecraft Data}

Let us now use the rule of thumb  in \eqref{eq:rule} with the empirically derived threshold value of $\mathcal{N}=12$ to assess the recovery of undersampled particle energization signatures in the analysis of spacecraft measurements from the 
\textit{Magnetospheric Multiscale} (\emph{MMS})\citep{Burch:2016} and \textit{Parker Solar Probe} (\emph{PSP})\citep{Fox:2016} missions.

Consider the case of the first \textit{in situ} observation of electron Landau damping that came from applying the field-particle correlation technique to \emph{MMS} observations of Earth's turbulent magnetosheath plasma in Chen \emph{et al.} (2019).\citep{Chen:2019} In energy interleave mode, the sampling of the 3V electron velocity distribution by the FPI instrument\citep{Pollock:2016} occurs on a cadence of $\Delta t=60$~ms, leading to a Nyquist frequency of $f_{Ny} \simeq 8.3$~Hz for the measurements. From Fig.~3 of Chen \emph{et al.} (2019),\citep{Chen:2019} the estimated onset of strong damping with $\gamma/\omega> 0.1$ occurs at $k_\perp \rho_i > 20$.  The transition from non-dispersive \Alfven waves to dispersive kinetic \Alfven waves occurs at $k_{\perp b} \rho_i \sim 1$, which is expected to be the perpendicular wavenumber associated with the break in the turbulent energy spectrum at ion scales.\citep{Howes:2008b}  In Fig.~5 of the same paper, the measured turbulent frequency spectrum of magnetic energy exhibits a break at $f_b \simeq 0.5$~Hz.  Assuming the Taylor Hypothesis to convert frequency measurements to wavenumber measurements,\citep{Taylor:1938} one may then estimate the frequency of kinetic \Alfven waves suffering significant collisionless damping as $f = ( k_\perp \rho_i /k_{\perp b} \rho_i) f_b \simeq 10$~Hz.  Therefore, these waves are slightly undersampled with $f/f_{Ny}=1.2$, and \eqref{eq:rule} predicts a minimum correlation interval of $\tau/T=7.2$, or a minimum dimensional correlation interval of $\tau \gtrsim 0.7$~s.  The full correlation interval used in the study is 70~s, so we expect any velocity-space signatures of energization to be statistically significant, consistent with the results of the phase randomization tests presented in that study.  Even for the shorter correlation intervals  of $\tau = 7$~s used in the timestack plot presented in their Fig.~2(b), the correlation interval used is significantly longer than the minimum suggested by \eqref{eq:rule}.

A follow up study of $20$ \emph{MMS} intervals in the turbulent magnetosheath ranging in length from $15$ to $150$ seconds also found FPC signatures of electron Landau damping in 95\% of the intervals examined.\citep{Afshari:2021} Assuming similar plasma parameters as for the electron Landau damping signature found in Chen \emph{et al.} (2019),\citep{Chen:2019} all of these intervals are still sufficiently long compared to the predicted minimum of $\tau \gtrsim 0.7$~s.

To seek signatures of electron Landau damping using  \textit{Parker Solar Probe} measurements in the inner heliosphere, we can follow the same general procedure as outlined above. The minimum cadence for sampling the 3V electron velocity distribution by the SPAN-E instrument\citep{Whittlesey:2020} is $\Delta t=0.2138$~s, leading to a  Nyquist frequency of $f_{Ny} \simeq 2.3$~Hz for the measurements. For Encounter 9 in August 2021 with a perihelion of 16~$R_\odot$, we take typical plasma parameters of $B \sim 100$~nT, $n_e \sim 140$~cm$^{-3}$, and $T_i \sim T_e \sim 50$~eV. Thus, we obtain dimensionless plasma parameters $\beta_i =0.28$ and $T_i/T_e \sim 1$ and a thermal ion Larmor radius of $\rho_i \simeq 10$~km.  The solar wind flow velocity is $v_{sw} \simeq 400$~km/s.  For these parameters, the doppler shifted frequency associated with the predicted ion-scale break in the magnetic energy spectrum at $k_{\perp b} \rho_i \sim 1$ has a frequency of $f_b \simeq 6.4$~Hz. For the plasma parameters specified, the linear Vlasov-Maxwell dispersion relation for kinetic \Alfven waves computed by the PLUME solver\citep{Klein:2015a} has an onset of strong damping with $\gamma/\omega \gtrsim 0.1$ at $k_\perp \rho_i \sim 5$. Thus, the predicted minimum frequency of kinetic \Alfven waves expected to experience significant electron Landau damping is $f = ( k_\perp \rho_i /k_{\perp b} \rho_i) f_b \simeq 32$~Hz.  The resulting undersampling factor is $f/f_{Ny}=14$, yielding a predicted minimum correlation interval of $\tau/T=84$, or a minimum dimensional correlation interval of $\tau \gtrsim 2.6$~s.  Therefore, it seems plausible that performing a field-particle correlation analysis over a correlation interval of $\tau \gtrsim 2.6$~s will be able to recover a velocity-space signature of electron Landau damping using \emph{PSP} measurements.

\section{ Conclusion }

Determining the mechanisms by which particles are energized through the dissipation of turbulence in weakly collisional space plasmas is a key goal of the heliophysics community, but observational identification of these energization mechanisms is hampered by the limited time resolution of the plasma particle measurements possible with existing spacecraft instrumentation. The field-particle correlation (FPC) technique is a new approach for the analysis of single-point spacecraft measurements to determine characteristic velocity-space signatures of the particle energization that can be used to identify specific energization mechanisms.  When taking frequency transforms of the time series of spacecraft measurements, the Nyquist frequency associated with the sampling cadence represents an upper limit preventing the recovery of information on the physical behavior at higher frequencies.  
The FPC technique, however,  does not utilize frequency transforms but instead simply computes the time-average of instantaneous energization rates. Therefore, we conjecture here that the physics of particle energization associated with fluctuations at frequencies above the Nyquist frequency ($f>f_{Ny}$) can be recovered because of the long-time averaging that is inherent to the FPC technique, even though those fluctuations are undersampled. Over a sufficiently long interval, even sparse discrete sampling will collect a representation of all phases of the large-amplitude oscillations that obscure the smaller-amplitude secular energy transfer, enabling cancellation of these oscillations upon averaging and recovery of the FPC energization signature, as illustrated in Fig.~\ref{fig:f_fNy}.

To demonstrate the recovery of undersampled energization signatures, we perform  a gyrokinetic simulation of a turbulent cascade in the kinetic \Alfven wave regime, $k_\perp \rho_i>1$, that has been evolved over nearly 70 domain-scale wave periods.
By applying the FPC technique to increasingly downsampled time series from this simulation, we show clearly that one can indeed recover the velocity-space signatures of electron Landau damping even when the KAWs are undersampled, answering the key science question of this study. This result is presented in Fig.~\ref{fig:taufreq_ao}, where the points in the unshaded region indicate resolved energization signatures with undersampling factors $f/f_{Ny}>1$. 

Furthermore, we construct an analytical model to determine empirically a rule of thumb, given by \eqref{eq:rule}, that provides guidance on the minimum duration of the correlation interval $\tau$ needed to recover a signature with an undersampling factor $f/f_{Ny}>1$.  We find that, effectively, one requires a minimum of $\mathcal{N}=12$  samples over the full correlation interval $\tau$ to recover an energization signature. Signatures recovered with fewer samples may not be statistically significant. If the particle energization is temporally intermittent (meaning the energization mechanism does not persist through the entire correlation interval, as illustrated in Fig.~\ref{fig:p14}), however, one may require a longer correlation interval to recover a reliable energization signature.

We first apply this rule of thumb to recently reported signatures of electron Landau damping in the Earth's turbulent magnetosheath using \emph{MMS} observations.$\citep{Chen:2019,Afshari:2021}$ We show that, even though those measurements are slightly undersampled with $f/f_{Ny}\gtrsim 1.2$, the correlation intervals used in all cases are significantly longer than the minimum recommended by \eqref{eq:rule}, providing additional evidence for their statistical significance.  

Next, we apply the rule of thumb to predict the minimum correlation interval needed to recover velocity-space signatures of electron Landau damping in the inner heliosphere using electron measurements from the \textit{Parker Solar Probe}  mission.  Under the plasma conditions in the inner heliosphere, we estimate that kinetic \Alfven waves which will experience significant electron Landau damping will have frequencies $f\gtrsim 32$~Hz. For the maximum sampling cadence of the electron velocity distributions ($\Delta t =0.2138$~s), this yields an undersampling factor $f/f_{Ny} \simeq 14$ and our rule of thumb predicts a minimum correlation interval of $\tau \gtrsim 2.6$~s. Therefore, a correlation interval of approximately $\tau \sim  10$~s should be sufficient to recover the velocity-space signature of electron Landau damping using \emph{PSP} observations. In conclusion, although the Nyquist frequency of the sampling of particle velocity distributions by \textit{Parker Solar Probe} is often lower than the frequencies of the turbulent fluctuations that are expected to energize the particles, our results suggest that the field-particle correlation technique can still be employed to address one of prime science questions of this flagship NASA mission---to determine how the solar corona is heated and the solar wind is accelerated.

\begin{acknowledgments}
S. A. H. was supported by NASA grant 80NSSC20K1509, A. J. M. was supported by NSF grant  AGS-1842561, and G. G. H. was supported by NASA grants 80NSSC18K0643 and 80NSSC18K1371. This work used the Extreme Science and Engineering Discovery Environment (XSEDE), which is supported by National Science Foundation grant number ACI-1548562, on Stampede2 at the Texas Advanced Computing Center through NSF XSEDE Award TG-PHY090084.
\end{acknowledgments}
The data that support the findings of this study are available from the corresponding author upon reasonable request.

\appendix 
\section{Automated Procedure for Determining Loss of Bipolar Signatures due to Downsampling}
\label{app:auto}
Here we describe in detail how we determine the downsampling factor $\delta$ at which we lose the bipolar signature identified in the full time-resolution set. 

\begin{enumerate}

\item For all probe points and downsampling versions, the approximate values of $v_\parallel$ for which both $C_{E_\parallel}(v_\parallel) = 0$ and $sgn(v_\parallel) [d C_{E_\parallel}(v_\parallel)/dv_\parallel] > 0$ are identified using an automated routine. This routine finds the slope ($m$) between a pair of consecutive parallel-velocity points ($v_{\parallel}^{(i)}, v_{\parallel}^{(i+1)}$), neglecting any cases for which $m \leq 10^{-3}$, and uses it to approximate the location of the zero point:
\begin{equation}
v_{\parallel} = v_\parallel^{(i)} - \frac{C_{E_\parallel}(v_\parallel^{(i)})}{m}.
\end{equation} 

\item At each probe point of the original dataset ($\delta = 0$), the zero-crossing locations identified in step 1 are visually compared with plots of $C_{E_\parallel}(v_\parallel)$ and $C_{E_\parallel}(v_\parallel, t; \tau =8.7)$. Any $v_\parallel$-locations that do not correspond to bipolar signatures of Landau damping are discarded.

\item  The confirmed signature locations are used in conjunction with the plots mentioned in step 2 to determine if the zero-crossings identified in the next downsampled dataset version correspond to these confirmed signatures or if a signature has been lost. We consider a signature to be last visible at downsampling factor $\delta$ if in dataset $\delta+1$ there is (a) no longer a zero-crossing or (b) the zero-crossing no longer corresponds clearly to a bipolar signature.

\item Step 3 is repeated, incrementing the downsampling exponent $\delta$ by 1,  until the confirmed signature locations from $\delta = 10$ have been compared with the identified zero-crossings from $\delta = 11$. 

\item The $v_\parallel$-location of the zero-crossing of a given bipolar signature varies somewhat as a dataset is downsampled. Thus, the average value across all appearances is taken and used to calculate the ratios of $f/f_{Ny}$ and $\tau/T$ using \eqref{eq:ffNy} and \eqref{eq:tauT}.

\end{enumerate}

\section{Analytic Model of Undersampled Signature Recovery }
\label{app:thumb}

An important physical consideration in interpreting the results of our downsampling tests shown in Fig.~\ref{fig:taufreq_ao} is that the waves suffering collisionless damping may not persist over the entire correlation interval. In other words, temporal intermittency of kinetic \Alfven wave fluctuations at a given perpendicular wavenumber may effectively reduce the sampling time taken while collisionless damping is occurring at the position of our probe.  In this case, the effective correlation interval for sampling the collisionless damping of a particular wave would be smaller than the full correlation interval $\tau$.  To remove this effect of temporal intermittency, in this appendix we build a simple mathematical model of recovering energization signatures from increasingly downsampled data.

The field-particle correlation technique uses the parallel correlation, given by \eqref{eq:cepar}, to compute the rate of energy transfer between the parallel electric field and the particles.  If we integrate this correlation over 3V velocity-space, we obtain the following
\begin{equation}
   \left(\frac{\partial W_s(\V{r}_0,t)}{\partial t}\right)_{E_\parallel} = 
   \left[ -q_s \int d^3\V{v}\frac{v_\parallel^2}{2} \frac{\partial f_s}{\partial v_\parallel} \right]E_\parallel  = j_{\parallel,s} E_\parallel 
\end{equation}
where $W_s(\V{r}_0,t)$ is the spatial energy density at the single point $\V{r}_0$. This result shows that the rate of change of spatial energy density of species $s$, due to the parallel electric field, is simply equal to the rate of work done by the parallel electric field on species $s$, $j_{\parallel,s} E_\parallel$. Each point in velocity space of the the field-particle correlation has the same mathematical form: the rate of energy transfer due to a small volume of particles in velocity-space is simply the product of the parallel current due to that small volume $\Delta j_{\parallel,s}$ multiplied by $E_\parallel$.  Therefore, we can build a simple mathematical model of how the FPC technique can recover the energy transfer rate and its velocity-space signature when the plasma waves are undersampled: we will model two scalar fields, $j_{\parallel,s}$ and  $E_\parallel$, and determine how well their product recovers the (well-sampled) energy transfer rate as each field is downsampled.

Our model specifies two well-sampled ($64$ samples per wave period) fields $j(t)$ and $E(t)$ oscillating at a frequency $\omega_0$ that represent the current and electric field fluctuations within the turbulent cascade at a particular perpendicular wavenumber. These model fields are constructed over 
many wave periods ($T \geq 2500$).  The number of discrete samples in each time series is then reduced by factors of two extending up to $2^{14}$, leading to a set of 15 datasets with downsampling exponents over the range $0 \le \delta \le 14$.  Thus, the sampling interval for a downsampled dataset is given by $\Delta t^{(\delta)} \equiv  \Delta t^{(0)} 2^\delta$, where the sampling interval of the original dataset is $\Delta t^{(0)}= 2 \pi /(64 \omega_0)$. The time-averaged energy transfer rate for these time series is defined by computing
$ \overline{\mathcal{E}}  = 1/N_{tot} \sum_{n=1}^{N_{tot}} j^{(0)}(t_n)E^{(0)}(t_n)$ using the original dataset, where $N_{tot}$ is the total number of discretely sampled times and the superscript denotes the downsampling exponent $\delta=0$.  Next, we define the 
energy transfer rate for a downsampled dataset over a "correlation interval" $\tau=N\Delta t^{(\delta)}$ by the cumulative time-average of the product of the downsampled time series, given by 
\begin{equation}
    \langle \mathcal{E}^{(\delta)} \rangle_{t_N} =
    \frac{1}{N} \sum_{n=1}^N j^{(\delta)}(t_n)E^{(\delta)}(t_n).
\end{equation}
The minimum correlation interval needed to accurately recover the energy transfer rate $ \overline{\mathcal{E}}$ is defined by determining the minimum correlation interval $\tau = t_N$ such that the error in the cumulative time-average $ \langle \mathcal{E}^{(\delta)} \rangle_{t_N}$ does not exceed 10\% as the averaging interval 
increases,
\begin{equation}
    \left|\frac{\langle \mathcal{E}^{(\delta)} \rangle_{t_N }-  \overline{\mathcal{E}}}{\overline{\mathcal{E}}} \right| \le 0.1
    \end{equation}
We may then compute the corresponding points in the $(f/f_{Ny},\tau/T)$ parameter space with this minimum value of $\tau$, using $f = 1/T = 2\pi \omega_0$ and $f_{Ny}=1/(2 \Delta t^{(\delta)})$.
 
Using two time series for $j(t)$ and $E(t)$ with a single oscillation frequency $\omega_0$ and a constant phase offset between the two fields is not a realistic model for the fluctuations in plasma turbulence at a particular perpendicular wavenumber. Therefore, to increase the relevance of the model, we make physically-motivated adjustments to factor in the transient nature of waves in turbulence. A decorrelation rate is incorporated into the frequency and small, random kicks are added to the amplitudes of $j$ and $E$ using a method based off of the oscillating Langevin antenna that is used to drive \T{AstroGK} turbulence in a physical manner.\citep{TenBarge:2014, Numata:2010} 

First, two complex waveforms, representing $j$ and $E$, are initialized with arbitrary initial real amplitudes $j_0$ and $E_0$ and a phase offset $\phi = 45$ degrees,
\begin{equation}
E_{n=0} = E_0,\ \ \ j_{n=0} = j_0 \cos{\phi} + i j_0 \sin{\phi}.
\end{equation}
 The complex coefficients $E_n$ and $j_n$ are advanced, following TenBarge \emph{et al.} (2014)\citep{TenBarge:2014}, by the expression
\begin{equation}
E_{n+1} = E_n e^{-i \omega_c \Delta t} + F_E \Delta t,
\label{eq:adv}
\end{equation}
along with the equivalent expression for $j$. 
In \eqref{eq:adv}, the complex frequency $\omega_c$ includes the real frequency $\omega_0$ plus a decorrelation rate $\gamma_0$, yielding, $\omega_c = \omega_0 + i \gamma_0$. We choose $\gamma_0/\omega_0 = -1/2\pi$ to model the decorrelation associated with the nonlinear energy transfer among different wave modes in the strong turbulent cascade. 
 The term that incorporates the random kicks to the amplitude is $F_E = \sigma_E u_n$, where 
\begin{equation}
 \sigma_E = E_0 \sqrt{12 \frac{| \gamma_0 |}{\Delta t^{(0)}}},
\end{equation}
as defined in TenBarge \emph{et al.} (2014)\citep{TenBarge:2014}, with the addition of a subscript to differentiate our usage for both $j$ and $E$. The random kicks $u_n$ are such that $Re\{u_n\} \in [-1/2, 1/2]$ and $Im\{u_n\} \in [-1/2, 1/2]$. Note that $j$ and $E$ are evolved individually before being combined into the product that is ultimately considered by the model ($\mathcal{E} = j_\parallel E_\parallel$),  but the same random kick $u_n$ is given to both waves at each time step to preserve their initial phase offset relative to each other.  

The effect of this particular oscillating Langevin model is to generate a time series that oscillates with an approximate real frequency $\omega_0$ and amplitude $E_0$ (or $j_0$), but with broadening of the frequency, phase, and amplitude of the oscillation.  An additional benefit of this approach, besides being a more realistic model of the turbulent fluctuations at a given wavenumber, is to prevent the uniform sampling in time $\Delta t^{(\delta)}$ from repeatedly sampling the same small set of wave phases---this prevents artifacts in the analysis we present below.

\begin{figure}[!h]
\includegraphics[width=0.49\textwidth, height=0.49\textwidth]{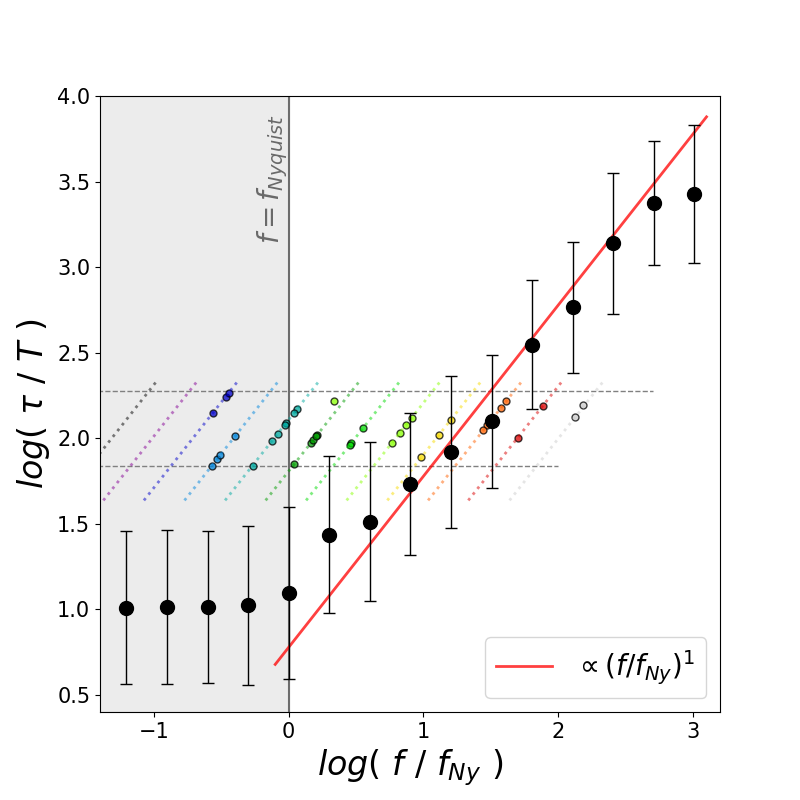}
\caption{\label{fig:model_phi45} The averaging interval $\tau/T$ required for a modeled wave to Nyquist frequency ratio $f/f_{Ny}$ to converge within 10\% of the true average value of the modeled energy transfer (average value in black, standard deviation given by the error bars), overlaid on the data from the energization signatures found in the simulation (color).}
\end{figure}

Using this physically motivated, idealized model of the fluctuations at a particular wavenumber in a turbulent plasma, we generate a statistical ensemble of $16$ independent datasets. Each of these datasets is then successively downsampled to create $15$ downsampled versions with downsampling exponents $0 \le \delta \le 14$. For each downsampling exponent, the cumulative time average $\langle \mathcal{E}^{(\delta)} \rangle_t$ is computed to determine the minimum correlation interval $\tau$ needed to obtain less than 10\% disagreement, and the mean  of the $\tau$ from the 16 independent datasets is kept.  This process of generating and analyzing $16$ independent datasets is repeated 32 times.  In  
Fig.~\ref{fig:model_phi45}, we plot the mean (black dots) and standard deviation (error bars) of the $\tau$ values from each of these 32 repetitions. In the  well-sampled region with $f/f_{Ny} < 1$ (shaded), an approximately  constant value of $\tau/T = 10$ is necessary to obtain an accurate estimate of the energy transfer rate.  For undersampled frequencies $f/f_{Ny}>1$, this plot shows clearly that the mean minimum correlation interval $\tau$ scales linearly with $f/f_{Ny}$ (red line).  

We also plot the data points resulting from analysis of the FPC signatures in the simulation, reproduced from Fig.~\ref{fig:taufreq_ao}, beneath this idealized model. These data are colored according to the downsampled dataset version $\delta$ in which each signature was last visible, and the colored dotted lines intersecting the simulation data points are exactly the bounding lines discussed in Sec.~\ref{sec:results}, except here we show each of the twelve.
 
To obtain an approximate rule of thumb for how long the correlation interval $\tau$ needs to be in order to recover a reliable energization signature, we look for a relation of the form
\begin{equation}
    \frac{\tau}{T} \gtrsim \frac{\mathcal{N}}{2} \frac{f}{f_{Ny}}.
    \label{eq:app_thumb}
\end{equation}
Taking the logarithm of this equation yields an equation of the form $y=mx+b$ where $y=\log(\tau/T)$, $x=f/f_{Ny}$, and the slope is $m=1$ since $f=1/T$. The coefficient $b = \log(\mathcal{N}/2)$ gives a threshold value needed to recover an energization signature.  Note that \eqref{eq:app_thumb} can be expressed 
$\tau/\Delta t= \mathcal{N}$, so given a sampling cadence $\Delta t$, the coefficient $\mathcal{N}$ corresponds to the number of samples taken within the correlation interval $\tau$. The red line in Fig.~\ref{fig:model_phi45} corresponds to 
$\mathcal{N}=12$, meaning that recovering a reliable energization signature requires at least 12 samples.

Note that we can interpret the simulation results plotted in  Fig.~\ref{fig:model_phi45} in the following way.  The points that lie at values $\tau/T$ above the rule of thumb are likely due to those waves not persisting in time over the entire correlation interval.  This temporal intermittency leads to an effectively smaller value of $\tau$ because the wave undergoing damping does not exist for the entire correlation interval.  The points below the rule of thumb are not likely to be statistically significant, so those signatures must be interpreted with extreme caution.


\end{document}